%% file: interference.tex
\documentclass[12pt,journal,final,onecolumn]{IEEEtran} 

\usepackage{amsfonts}
\usepackage[cmex10]{amsmath}
\usepackage{amssymb}
\usepackage{amsthm}
\usepackage{mathrsfs}
\usepackage{cite}
\usepackage{graphicx}
\usepackage[caption=false]{subfig}
\usepackage{color,colortbl}

\usepackage{etoolbox}
\makeatletter
\patchcmd{\@makecaption}
  {\scshape}
  {}
  {}
  {}
\makeatother

\graphicspath{ {./figures/} }

\input{niesen.def}
\renewcommand{\epsilon}{\varepsilon}
\newcommand{\myfigsscale}{.8}
\newcommand{\mygainsscale}{.9}

\definecolor{user1}{rgb}{0.8,1,1}
\definecolor{user2}{rgb}{1,0.8,1}

\begin{document}
\title{Degrees of Freedom of\\Cache-Aided Wireless Interference Networks}

\author{Jad~Hachem, Urs~Niesen, and~Suhas~Diggavi
\thanks{J. Hachem and S. Diggavi are with the Department of Electrical Engineering,
University of California, Los Angeles.
Emails: \textsf{jadhachem@ucla.edu}, \textsf{suhas@ee.ucla.edu}.}%
\thanks{U. Niesen is with the Qualcomm NJ Research Center, Bridgewater, NJ.
Email: \textsf{urs.niesen@ieee.org}.}%
\thanks{This work was supported in part by NSF grant \#1423271.}%
\thanks{This paper was presented in part at the 2016 IEEE International Symposium on Information Theory in Barcelona, Spain.}}

\maketitle

\begin{abstract}
\input{input_files/abstract.tex}

\end{abstract}


\section{Introduction}
\label{sec:intro}
\input{input_files/intro.tex}

\section{Problem Setting}
\label{sec:setup}
\input{input_files/setup.tex}

\section{Main Results}
\label{sec:results}
\input{input_files/results.tex}

\section{Separation Architecture}
\label{sec:separation}
\input{input_files/separation.tex}

\section{The Multiple Multicast X-Channel}
\label{sec:alignment}
\input{input_files/alignment.tex}

\section{Order-Optimality of the Separation Architecture}
\label{sec:converse}
\input{input_files/converse.tex}

\section{An Alternative Separation Strategy}
\label{sec:2x2}
\input{input_files/2x2.tex}

\section{Discussion}
\label{sec:discussion}
\input{input_files/discussion.tex}

\appendices
\section{Special Case: Small Number of Files}
\label{app:small-n}
\input{input_files/small-n.tex}

\section{Proof of Lemma~\ref{lemma:phy-achievability}}
\label{app:alignment-details}
\input{input_files/alignment-details.tex}

\section{Detailed Converse Proof of Theorem~\ref{thm:dof}}
\label{app:converse}
\input{input_files/converse-detailed.tex}

\section{Communication Problem Outer Bounds (Converse Proof of Theorem~\ref{thm:phy})}
\label{app:phy-converse}
\input{input_files/phy-converse.tex}

\section{Lemmas from \cite{cadambe2009}}
\label{app:cadambe-lemmas}
\input{input_files/cadambe-lemmas.tex}

\section{Proof of Theorem~\ref{thm:2x2}}
\label{app:2x2}
\input{input_files/2x2-proofs.tex}

\bibliographystyle{IEEEtran}
\bibliography{journal_abbr,caching,xchannel}


\end{document}

%% file: input_files/abstract.tex
We study the role of caches in wireless interference networks.
We focus on content caching and delivery across a Gaussian interference network, where both transmitters and receivers are equipped with caches.
We provide a constant-factor approximation of the system's degrees of freedom (DoF), for arbitrary number of transmitters, number of receivers, content library size, receiver cache size, and transmitter cache size (as long as the transmitters combined can store the entire content library among them).
We demonstrate approximate optimality with respect to information-theoretic bounds that do not impose any restrictions on the caching and delivery strategies.
Our characterization reveals three key insights.
First, the approximate DoF is achieved using a strategy that separates the physical and network layers.
This separation architecture is thus approximately optimal.
Second, we show that increasing transmitter cache memory beyond what is needed to exactly store the entire library between all transmitters does not provide more than a constant-factor benefit to the DoF.
A consequence is that transmit zero-forcing is not needed for approximate optimality.
Third, we derive an interesting trade-off between the receiver memory and the number of transmitters needed for approximately maximal performance.
In particular, if each receiver can store a constant fraction of the content library, then only a constant number of transmitters are needed.
Our solution to the caching problem requires formulating and solving a new communication problem, the symmetric multiple multicast X-channel, for which we provide an exact DoF characterization.

%% file: input_files/intro.tex
Traditional communication networks focus on establishing a reliable connection between two fixed network nodes and are therefore connection centric.
With the recent explosion in multimedia content, network usage has undergone a significant shift: users now want access to some specific content, regardless of its location in the network.
Consequently, network architectures are shifting towards being content centric.
These content-centric architectures make heavy use of in-network caching and, in doing so, redesign the protocol stack from the network layer upwards~\cite{jacobson2012}.

A natural question to ask is how the availability of in-network caches can be combined with the wireless physical layer and specifically with two fundamental properties of wireless communication: the broadcast and the superposition of transmitted signals.
Recent work in the information theory literature has demonstrated that this combination can yield significant benefits.
This information-theoretic approach to caching was introduced in the context of the noiseless broadcast channel in \cite{maddah-ali2012}, where it was shown that significant performance gains can be obtained using cache memories at the \emph{receivers}.
In \cite{maddah-ali2015interference}, the noiseless broadcast setting was extended to the interference channel, which is the simplest multiple-unicast wireless topology capturing both broadcast and superposition.
The authors presented an achievable scheme showing performance gains using cache memories at the \emph{transmitters}.

\begin{figure}
\centering
\includegraphics[scale=\myfigsscale]{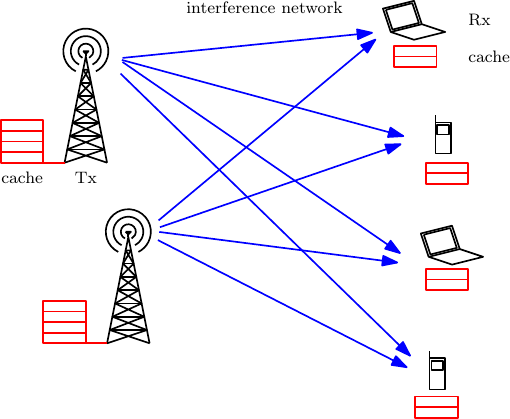}
\caption{Caching in a wireless interference network.
Caches (in red) are placed at all network nodes.}
\label{fig:setup-simple}
\end{figure}

In this paper, we continue the study of the cache-aided wireless interference network, but we allow for caches at both the \emph{transmitters and receivers} as shown in \figurename~\ref{fig:setup-simple}.
Our main result (Theorem~\ref{thm:dof}, Section~\ref{sec:results}) is a complete constant-factor approximation of the degrees of freedom (DoF) of this network.
The result is general, in that it holds for any number of transmitters and receivers, size of content library, transmitter cache size (large enough to collectively hold the entire content library), and receiver cache size.
Moreover, our converse holds for arbitrary caching and transmission functions, and imposes no restrictions as done in prior work.

Several architectural and design insights emerge from this degrees-of-freedom approximation.
\begin{enumerate}
\item
Our achievable scheme introduces a novel separation of the physical and network layers, thus redesigning the protocol stack from the network layer downwards.
From the order-wise matching converse, we hence see that this separation is approximately optimal.

\item
Once the transmitter caches are large enough to collectively hold the entire content library, increasing the transmitter memory further can lead to at most a constant-factor improvement in the system's degrees of freedom.
In particular, and perhaps surprisingly, this implies that transmit zero-forcing is not needed for approximately optimal performance.

\item
There is a trade-off between the number of transmitters needed for (approximately) maximal system performance and the amount of receiver cache memory.
As the receiver memory increases, the required number of transmitters decreases, down to a constant when the memory is a constant fraction of the entire content library.
\end{enumerate}

There are three seemingly natural network-layer abstractions for this problem.
The first network-layer abstraction treats the physical layer as a standard interference channel and transforms it into non-interacting bit pipes between disjoint transmitter-receiver pairs.
This approach is inefficient.
The second network-layer abstraction treats the physical layer as an X-channel and transforms it into non-interacting bit pipes between each transmitter and each receiver.
The third network-layer abstraction treats the physical layer as multiple broadcast channels: it creates a broadcast link from each transmitter to all receivers.
The last two approaches turn out to be approximately optimal in special circumstances: the second when the receivers have no memory, and the third when they have enough memory to each store almost all the content library.
In this paper, we propose a network-layer abstraction that creates X-channel \emph{multicast} bit pipes, each sent by a transmitter and intended for a subset of receivers whose size depends on the receiver memory.
This abstraction generalizes the above two approaches, and we show that it is in fact order-optimal for all values of receiver memory.

Our solution to this problem requires solving a new communication problem at the physical layer that arises from the proposed separation architecture.
This problem generalizes the X-channel setting studied in \cite{cadambe2009} by considering multiple \emph{multicast} messages instead of just unicast.
We focus on the symmetric case and provide a complete and exact DoF characterization of this symmetric multiple multicast X-channel problem, by proposing a strategy based on interference alignment and proving its optimality (see Theorem~\ref{thm:phy}, Section~\ref{sec:separation}).

\subsection*{Related Work}

Content caching has a rich history and has been studied extensively,
see for example \cite{Wessels:2001} and references therein.
Recent interest in content caching is motivated by Video-on-Demand
systems for which efficient content placement and delivery schemes have
been proposed in \cite{korupolu1999, Borst:2010, Tan:2013,
llorca2013}. The impact of content popularity distributions on caching
schemes has also been widely investigated, see for
example \cite{Wolman99, breslau1999web, Applegate:2010}.
Most of the literature has focused on wired networks, and the solutions
there do not carry directly to wireless networks.

The information-theoretic framework for coded caching was introduced in
\cite{maddah-ali2012} in the context of the deterministic broadcast
channel.  This has been extended to online caching systems \cite{PMN13},
systems with delay-sensitive content \cite{niesen2014},
heterogeneous cache sizes \cite{wang2015fundamental},
unequal file sizes \cite{zhang2015filesize}, and improved converse
arguments \cite{ghasemi2015improved, sengupta2015improved}. Content
caching and delivery in device-to-device networks, multi-server
topologies, and heterogeneous wireless networks have been studied in
\cite{ji2013wireless,shariatpanahi2015multi,FemtoCaching,MLpopularityaccess}.
This framework was also applied to hierarchical (tree) topologies in
\cite{Hcodedcaching}, and to non-uniform content popularities in
\cite{niesen2013,Zcodedcaching,ZhangArbitrary,ji2015random,MLpopularityaccess}.
Other related work includes \cite{Gitzenis13}, which
derives scaling laws for content replication in multihop wireless
networks, and \cite{Ioannidis:2010}, which explores distributed caching in
mobile networks using device-to-device
communications.
The benefit of coded caching when the caches are randomly distributed was studied in \cite{AltmanAG13}, and the benefits of adaptive content placement using knowledge of user requests were explored in~\cite{YangH13}.

More recently, this information-theoretic framework for coded caching has been extended in \cite{maddah-ali2015interference} to interference channels with caches at only the transmitters, focusing on three transmitters and three receivers.
The setting was extended in \cite{sengupta2016} to arbitrary numbers of transmitters and receivers and included a rate-limited fronthaul.
Interference channels with caches \emph{both} at transmitters and at receivers were considered in \cite{naderializadeh2016,xu2016,layeredcaching}, all of which have a setup similar to the one in this paper.
However, each of these three works has some restrictions on the setup.
The authors in \cite{naderializadeh2016} focus on one-shot linear schemes, while \cite{xu2016} prohibits inter-file coding during placement and limits the number of receivers to three.
Our prior work \cite{layeredcaching} studies the same setup but with only two transmitters and two receivers.
The work in this paper differs from those above in that it considers an arbitrary number of transmitters and receivers and proves order-optimality using outer bounds that assume \emph{no restrictions} on the scheme.

Because we have overlapping results with \cite{xu2016,xu2016-arxiv}, we here give a timeline of the results as published on arXiv.
The first version of \cite{xu2016-arxiv} was placed on arXiv in May 2016 and discussed a similar setup as in this paper but with only two or three receivers, as well as an outer bound that prohibits inter-file coding during placement.
It is similar to the version \cite{xu2016} published in ISIT July 2016.
In June 2016, we posted an initial draft of this paper on arXiv \cite{dof-caching-v1} with all the results given in this paper: a general setup with an arbitrary number of transmitters and receivers, a separation-based strategy, and general information-theoretic outer bounds that pose no restrictions on the strategy and proves approximate optimality of our strategy.
To the best of our knowledge, these are the first approximate optimality results for a cache-aided interference channel with caches at both the transmitters and receivers.
In March 2017, another version of \cite{xu2016-arxiv} was posted on arXiv that included our general result (approximate degrees of freedom in the general case), which had appeared in \cite{dof-caching-v1}.
However, their approximate optimality result and proof were almost identical to ours \cite{dof-caching-v1}.
It also included a scheme that can achieve a constant-factor improvement over ours in the case with three receivers, and an extension of our scheme to a regime that we exclude in this paper (total transmitter memory less than the size of the content library).

\subsection*{Organization}

The remainder of this paper is organized as follows.
Section~\ref{sec:setup} introduces the problem setting and establishes notation.
Section~\ref{sec:results} states the paper's main results.
Section~\ref{sec:separation} presents the separation architecture in detail; Section~\ref{sec:alignment} gives the interference alignment strategy used at the physical layer.
Section~\ref{sec:converse} proves the order-optimality of our strategy.
Section~\ref{sec:2x2} explores an interesting variant of the separation architecture.
Section~\ref{sec:discussion} discusses extensions to the problem as well as relation to some works in the literature.
We defer additional proofs to the appendices.

%% file: input_files/setup.tex
\begin{figure}
\centering
\includegraphics[scale=\myfigsscale]{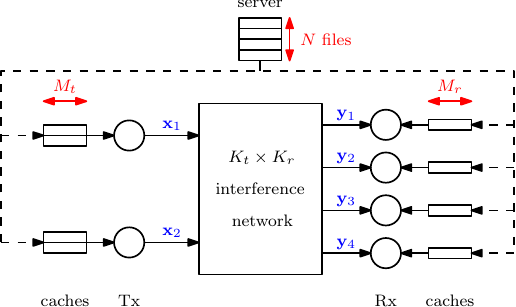}
\caption{The caching problem, with $K_t=2$ transmitters and $K_r=4$ receivers.
    The server holds a content library of $N$ files.
	Information about these files is placed in the transmitter caches of size $M_t$ and in the receiver caches of size $M_r$ during a placement phase (indicated by dashed lines).
During the subequent delivery phase (indicated by solid lines), each user requests one file, and all the requested files have to be delivered over the interference network.}
\label{fig:setup2x4}
\end{figure}

A content library contains $N$ files $W_1,\ldots,W_N$ of size $F$ bits each.
A total of $K_r$ users will each request one of these files, which must be transmitted across a $K_t\times K_r$ time-varying Gaussian interference channel whose receivers are the system's users.
We will hence use the terms ``receiver'' and ``user'' interchangeably.
Our goal is to reliably transmit these files to the users with the help of caches at both the transmitters and the receivers.

\begin{example}
\label{eg:2x4}
The setup is depicted in \figurename~\ref{fig:setup2x4} for the case with $K_t=2$ transmitters, $K_r=4$ receivers, and $N=4$ files in the content library.
We will use this setting as a running example throughout the paper.
\end{example}

The system operates in two phases, a \emph{placement phase} and a \emph{delivery phase}.
In the placement phase, the transmitter and receiver caches are filled as an arbitrary function of the content library.
The transmitter caches are able to store $M_tF$ bits; the receiver caches are able to store $M_rF$ bits.
We refer to $M_t$ and $M_r$ as the transmitter and receiver cache sizes, respectively.
Other than the memory constraints, we impose no restrictions on the caching functions (in particular, we allow the caches to arbitrarily code across files).
In this paper, we consider all values of $M_r\ge0$, but we restrict ourselves to the case where the transmitter caches can collectively store the entire content library,%
\footnote{To achieve any positive DoF, the minimum requirement is that $K_tM_t+M_r\ge N$, i.e., that all the transmitter caches and any single receiver cache can collectively store the entire content library.
	We impose the slightly stronger requirement $K_tM_t \geq N$ since we believe that it is the regime of most practical interest, and since it simplifies the analysis.}
i.e.,
\begin{equation}
\label{eq:mt-condition}
M_t\ge N/K_t.
\end{equation}

The delivery phase takes place after the placement phase is completed.
In the beginning of the delivery phase, each user requests one of the $N$ files.
We denote by $\mathbf{u}=(u_1,\ldots,u_{K_r})$ the vector of user demands, such that user $i$ requests file $W_{u_i}$.
These requests are communicated to the transmitters, and each transmitter $j$ responds by sending a codeword $\mathbf{x}_j=(x_j(1),\ldots,x_j(T))$ of block length $T$ into the interference channel.
We impose a power constraint over every channel input $\mathbf{x}_j$,
\[
    \frac1T \norm{\mathbf{x}_j}^2 \le \SNR,\quad\forall j=1,\ldots,K_t.
\]
Note that each transmitter only has access to its own cache, so that $\mathbf{x}_j$ only depends on the contents of transmitter $j$'s cache and the user requests $\mathbf{u}$.
We impose no other constraint on the channel coding function (in particular, we explicitly allow for coding across time using potentially nonlinear schemes).

Receiver $i$ observes a noisy linear combination of all the transmitted codewords,
\[
y_i(\tau) \defeq \sum_{j=1}^{K_t} h_{ij}(\tau) x_j(\tau) + z_i(\tau),
\]
for all time instants $\tau=1,\ldots,T$, where the $z_i(\tau)$'s are independent identically distributed (iid) unit-variance additive Gaussian noise, and $h_{ij}(\tau)$ are independent time-varying random channel coefficients obeying some continuous probability distribution.
We can rewrite the channel outputs in vector form as
\begin{equation}
\label{eq:yi}
\mathbf{y}_i = \sum_{j=1}^{K_t} \mathbf{H}_{ij} \mathbf{x}_j + \mathbf{z}_i,
\end{equation}
where $\mathbf{H}_{ij}$ is a diagonal matrix representing the channel coefficients over the block length $T$.

For fixed values of $M_t$, $M_r$, and $\SNR$, we say that a transmission rate $R=F/T$ is achievable if there exists a coding scheme such that all the users can decode their requested files with vanishing error probability.
More formally, $R$ is \emph{achievable for demand vector $\mathbf{u}$} if
\[
    \max_{i\in\{1,\ldots,K_r\}}\Pp\bigl( \hat W_i \neq W_{u_i} \bigr)
\to 0 \text{ as } T\to\infty,
\]
where $\hat W_i$ indicates the reconstruction of file $W_{u_i}$ by user $i$.
Note that $R$ is fixed as $T$, and hence $F$, go to infinity.
We say $R$ is \emph{achievable} if it is achievable for all demand vectors $\mathbf{u}$.

We define the optimal transmission rate $R^\star(\SNR)$ as the supremum of all achievable rates for a given $\SNR$ (and number of files, cache sizes, and number of transmitters/receivers).
In the remainder of this paper we will focus on the degrees of freedom (DoF) defined as
\begin{equation}
\label{eq:dof-def}
\DoF \defeq \lim_{\SNR\to\infty} \frac{R^\star(\SNR)}{\frac12\log\SNR}.
\end{equation}
While the DoF is useful for presenting and interpreting the main results in the next section, we will also often work with its reciprocal $1/\DoF$ because it is a convex function of $(M_t,M_r)$.

%% file: input_files/results.tex
The main result of this paper is a complete constant-factor approximation of the DoF for the cache-aided wireless interference network.
In order to state the result, we define the function $d(N,K_t,K_r,M_t,M_r)$---which we will sometimes write $d(\cdot)$ for simplicity---through
\begin{equation}
\label{eq:achievable}
\frac{1}{d(\cdot)}
\triangleq
\frac{K_t-1+\min\left\{\frac{K_r}{\kappa+1},N\right\}}{K_t}
\cdot \left( 1 - \frac{\kappa}{K_r} \right),
\end{equation}
for any $N$, $K_t$, $K_r$, $M_t$, and $M_r=\kappa N/K_r$ with $\kappa\in\{0,1,\ldots,K_r\}$, and the lower convex envelope of these points for all other $M_r\in[0,N]$.

\begin{figure}
\centering
\includegraphics[scale=\myfigsscale]{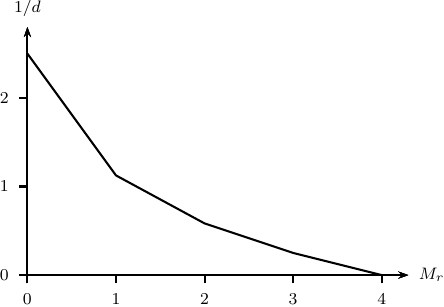}
\caption{Approximate reciprocal DoF of the $2\times 4$ cache-aided interference network with $4$ files, introduced in Example~\ref{eg:2x4}, as a function of receiver cache size $M_r$, for any $M_t\ge N/K_t$.}
\label{fig:dof24}
\end{figure}

\begin{theorem}
\label{thm:dof}
The degrees of freedom $\DoF$ of the $K_t\times K_r$ cache-aided interference network with $N$ files, transmitter cache size $M_t\in [N/K_t, N]$, and receiver cache size $M_r\in[0,N]$ satisfies
\[
d(N,K_t,K_r,M_t,M_r) \le \DoF \le 13.5 \cdot d(N,K_t,K_r,M_t,M_r).
\]
\end{theorem}
The approximate (reciprocal) DoF is illustrated in \figurename~\ref{fig:dof24} for the setup in Example~\ref{eg:2x4}.

In terms of the rate $R^\star(\SNR)$ of the system, Theorem~\ref{thm:dof} can be interpreted using \eqref{eq:dof-def} as
\begin{IEEEeqnarray*}{rCl}
&& d(\cdot) \cdot \frac12\log\SNR - o\left( \log\SNR \right)\\
&\le& R^\star(\SNR)\\
&\le& 13.5 \cdot d(\cdot) \cdot \frac12\log\SNR + o\left( \log\SNR \right),
\end{IEEEeqnarray*}
when $\SNR$ grows, where we have again used $d(\cdot)$ instead of $d(N,K_t,K_r,M_t,M_r)$ for simplicity.

The constant $13.5$ in Theorem~\ref{thm:dof} is the result of some loosening of inequalities in order to simplify the analysis.
We numerically observe that the multiplicative gap does not exceed $4.16$ for $N,K_t,K_r\le100$.

The coding scheme achieving the lower bound on $\DoF$ in Theorem~\ref{thm:dof} uses separate network and physical layers.
The two layers interface using a set of multicast messages from each transmitter to many subsets of receivers.
At the physical layer, an interference alignment scheme (generalizing the scheme from \cite{cadambe2009}) delivers these messages across the interference channel with vanishing error probability and at optimal degrees of freedom.
At the network layer, a caching and delivery strategy generalizing the one in \cite{maddah-ali2012} is used to deliver the requested content to the users, utilizing the non-interacting error-free multicast bit pipes created by the physical layer.
The matching upper bound in Theorem~\ref{thm:dof} shows that this separation approach is without loss of order optimality.
This separation architecture is described in more detail in Section~\ref{sec:separation}.

In order to better understand the behavior of the system, we decompose the approximation of the sum degrees of freedom $K_r\DoF$ provided by Theorem~\ref{thm:dof} into three components, or gains.%
\footnote{Note that this decomposition arises from our interpretation of our approximately optimal strategy described in Section~\ref{sec:separation}.}
These are: an interference alignment (IA) gain $g^\mathrm{IA}$, a local caching gain $g^\mathrm{LC}$, and a global caching gain $g^\mathrm{GC}$, forming
\begin{IEEEeqnarray*}{rCl}
\IEEEeqnarraymulticol{3}{l}{K_r \DoF}\\
\ &\approx& K_r d(N,K_t,K_r,M_t,M_r)\\
&\overset{(a)}{\approx}&
\underbrace{ \frac{K_tK_r}{K_t+K_r-1} }_{g^\mathrm{IA}}
\cdot \underbrace{ \frac{1}{1 - \frac{M_r}{N}} }_{g^\mathrm{LC}}
\cdot \underbrace{
    \frac{K_rM_r/N+1}{\frac{M_r}{N}\bigl(\frac{1}{K_r}+\frac{1}{K_t-1}\bigr)^{-1}+1}
}_{g^\mathrm{GC}}.
\IEEEyesnumber\label{eq:dof-gains}
\end{IEEEeqnarray*}
Note that $(a)$ holds with exact equality when $K_rM_r/N$ is an integer.
We point out that, for ease of presentation, this decomposition is written for the case when the first term achieves the minimum in \eqref{eq:achievable}, i.e., $K_r/(\kappa+1)\le N$.
This includes the most relevant case when the content library $N$ is larger than the number of receivers $K_r$.
In fact, we focus on this case in most of the main body of the paper, particularly regarding the achievability and some of the intuition.
A detailed discussion of the case $K_r/(\kappa+1)>N$, including a decomposition similar to \eqref{eq:dof-gains}, is given in Appendix~\ref{app:small-n}.

The term $g^\mathrm{IA}$ is the degrees of freedom achieved by communication using interference alignment and is the same as in the unicast X-channel problem \cite{cadambe2009}.
It is the only gain present when the receiver cache size is zero.
In other words, it is the baseline degrees of freedom without caching (see for example \figurename~\ref{fig:dof24} when $M_r=0$).

When the receiver cache size is non-zero, we get two improvements, in analogy to the two gains described in the broadcast caching setup in \cite{maddah-ali2012}.
The local caching gain reflects that each user already has some information about the requested file locally in its cache.
Hence, $g^\mathrm{LC}$ is a function of $M_r/N$, the fraction of each file stored in a single receiver cache.
On the other hand, the global gain derives from the coding opportunities created by storing different content at different users, and from the multicast links created to serve coded information useful to many users at once.
This gain depends on the total amount of receiver memory, as is reflected by the $K_rM_r/N$ term in the numerator of $g^\mathrm{GC}$.

It is interesting to see how each of these gains scales with the various system parameters $K_t$, $K_r$, $M_t/N$, and $M_r/N$.
In order to separate the different gains, we work with the logarithm
\[
\log K_r\DoF
\approx \log g^\mathrm{IA}
+ \log g^\mathrm{LC}
+ \log g^\mathrm{GC}
\]
of the sum degrees of freedom.
By varying the different parameters, we can plot how both the sum DoF and its individual components evolve.

\subsubsection{Scaling with transmitter memory $M_t$}

Notice in Theorem~\ref{thm:dof} that the DoF approximation does not involve the transmitter memory $M_t$.
Thus, once $M_t = N/K_t$, just enough to store the entire content library between all transmitters, any increase in the transmit memory will only lead to at most a constant-factor improvement in the DoF.

The strategy used to achieve the lower bound in Theorem~\ref{thm:dof} (see Section~\ref{sec:separation} for details) stores uncoded \emph{nonoverlapping} file parts in each transmit cache.
This is done regardless of the transmitter memory $M_t$ and the receiver memory $M_r$.
Since this is an order-optimal strategy, we conclude that the transmitters do not need to have any shared information.
Consequently, and perhaps surprisingly, transmit zero-forcing is not needed for order-optimality and cannot provide more than a constant-factor DoF gain.
Moreover, given that the value $13.5$ of the constant gap is close to and was obtained using similar arguments to the value of $12$ derived in \cite{maddah-ali2012} for the error-free broadcast case, we conjecture that most of the improvements on the constant would not come from sharing information among transmitters or from any transmit zero-forcing, but rather from tighter converse arguments.

\subsubsection{Scaling with receiver memory $M_r$}

\begin{figure}
\centering
\includegraphics[scale=\mygainsscale]{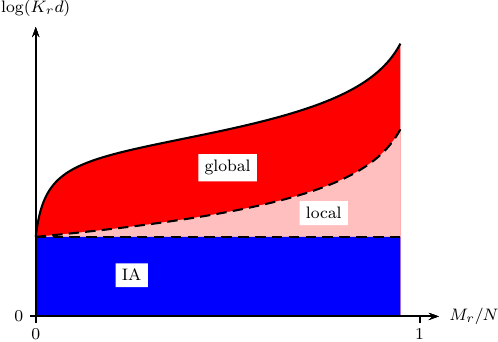}
\caption{DoF gains as a function of receiver cache size characterized by $M_r/N$.}
\label{fig:gains-mr}
\end{figure}

\figurename~\ref{fig:gains-mr} depicts the decomposition of the approximate sum degrees of freedom $K_r d \approx K_r\DoF$ as a function of the receiver cache size $M_r$.
As expected, the interference alignment gain $g^\mathrm{IA}$ does not depend on the receiver cache size and is hence constant.
The local caching gain $g^\mathrm{LC}$ increases slowly with $M_r$ and becomes relevant whenever each receiver can cache a significant fraction of the content library, say $M_r/N \geq 0.5$.
The global caching gain $g^\mathrm{GC}$ increases much more quickly and is relevant whenever the cumulative receiver cache size is large, say $K_r M_r /N \geq 1$.

\subsubsection{Scaling with number of receivers $K_r$}

\begin{figure}
\centering
\includegraphics[scale=\mygainsscale]{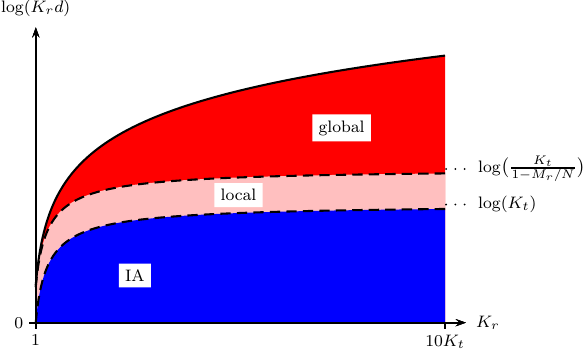}
\caption{DoF gains as a function of number of receivers $K_r$.}
\label{fig:gains-kr}
\end{figure}

\figurename~\ref{fig:gains-kr} depicts the decomposition of the approximate sum degrees of freedom $K_r d \approx K_r\DoF$ as a function of the number of receivers $K_r$.
The local caching gain $g^\mathrm{LC}$ is not a function of $K_r$ and is hence constant as expected.
In the limit as $K_r\to\infty$, the interference alignment gain $g^\mathrm{IA}$ converges to $K_t$.
The global caching gain $g^\mathrm{GC}$, on the other hand, behaves as
\begin{equation*}
    g^\mathrm{GC} \approx \frac{K_rM_r/N+1}{(K_t-1)M_r/N+1}
\end{equation*}
for large $K_r$.
In particular, unlike the other two gains, the global gain does not converge to a limit and scales linearly with the number of receivers.
Thus, for systems with larger number of receivers, the global caching gain becomes dominant.

\subsubsection{Scaling with number of transmitters $K_t$}
\label{sec:scaling-kt}

As the number of receivers $K_r$ or the receive memory $M_r$ increase, the sum DoF grows arbitrarily large.
The same is not true as the number of transmitters $K_t$ increases.
In fact, as $K_t\to\infty$, we find that $g^\mathrm{IA}\to K_r$, $g^\mathrm{GC}\to1$, and the sum DoF converges to
\begin{equation}
\label{eq:dof-lim-kt}
    \lim_{K_t\to\infty}K_r\DoF \approx \frac{K_r}{1-M_r/N}.
\end{equation}
This is not surprising, since, with a large number of transmitters, interference alignment effectively creates $K_r$ orthogonal links from each transmitter to the receivers, each of DoF approaching $1$.
With the absence of multicast due to these orthogonal links, the global caching gain vanishes and the only caching gain left is the local one.

\begin{figure}[htbp]
    \centering
    \subfloat[Small $M_r/N$.]{\label{fig:gains-kt3}\includegraphics[scale=\mygainsscale]{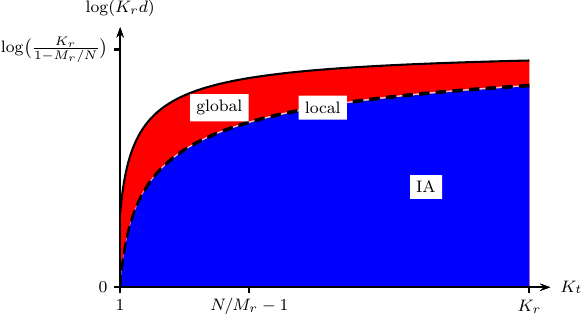}}\\
    \subfloat[Moderate $M_r/N$.]{\label{fig:gains-kt2}\includegraphics[scale=\mygainsscale]{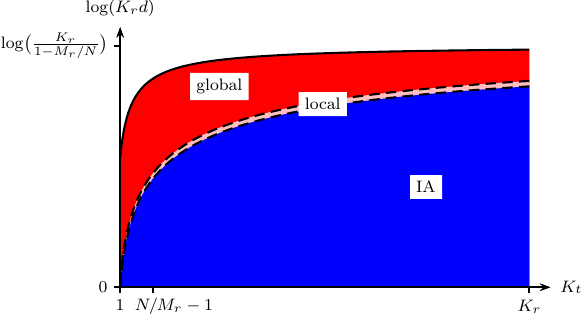}}\\
    \subfloat[Large $M_r/N$.]{\label{fig:gains-kt1}\includegraphics[scale=\mygainsscale]{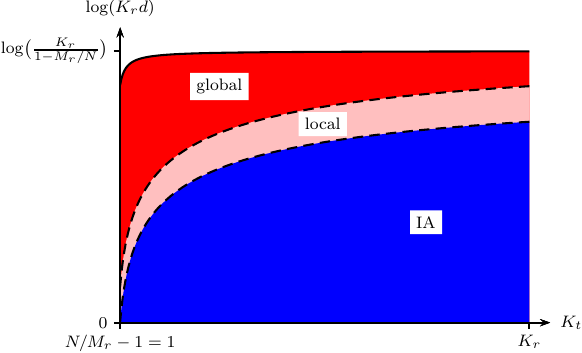}}

    \caption{DoF gains as a function of number of transmitters $K_t$ for various regimes of receiver cache size (characterized by $M_r/N$).}
    \label{fig:gains-kt}
\end{figure}

An interesting question then is how large $K_t$ has to be for the DoF to approach the limit in \eqref{eq:dof-lim-kt}.
Specifically, for what values of $K_t$ does the sum DoF become $\Theta\bigl(K_r/(1-M_r/N)\bigr)$?
When the receiver cache memory is small, specifically $M_r<N/K_r$, the number of transmitters $K_t$ must be of the order of $K_r$ (see \figurename~\ref{fig:gains-kt3}).
However, as $M_r$ increases, we find that a smaller number of transmitters is needed to achieve the same DoF (see Figs.~\ref{fig:gains-kt2} and~\ref{fig:gains-kt1}).
In general, the limiting value is reached (within a constant) when $K_t=\Omega(N/M_r-1)$.%
\footnote{This comes from being able to write $K_r\DoF\approx\frac{K_t(\kappa+1)}{K_t(\kappa+1)+K_r-\kappa-1} \cdot \frac{K_r}{1-M_r/N}$, where $\kappa=K_rM_r/N$.
The first factor is a constant when $K_t(\kappa+1)=\Omega(K_r-\kappa-1)$, which leads to $K_t=\Omega(\frac{K_r}{K_rM_r/N+1}-1)$.
When $K_r$ is large, this behavior becomes $K_t=\Omega(\frac{N}{M_r}-1)$.}
In particular, if the receiver caches can store a constant fraction of the content library, then we only need a constant number of transmitters to achieve maximal benefits, up to a multiplicative constant.
There is thus a trade-off between the number of transmitters $K_t$ and the amount of receiver cache memory $M_r$ required for maximal system performance (up to the local caching gain): the larger the receiver memory, the fewer the required transmitters.

While the separation architecture discussed above (on which we focus in most of this paper) is order optimal, one can still make some strict improvements, albeit no more than a constant factor, by choosing a different separation architecture.
In Section~\ref{sec:2x2}, we present an alternative separation architecture for the case $K_t=K_r=N=2$ that creates \emph{interacting} error-free bit pipes as the physical-layer abstraction.
This architecture can achieve a strictly higher DoF than Theorem~\ref{thm:dof} in some regimes.

%% file: input_files/separation.tex
\begin{figure}
\centering
\subfloat[Physical-layer view.]{%
	\label{fig:separation-phy}
	\includegraphics[scale=\myfigsscale]{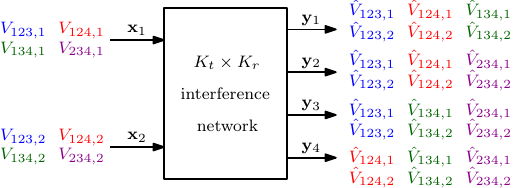}
}
\\
\subfloat[Network-layer view.]{%
	\label{fig:separation-net}
	\includegraphics[scale=\myfigsscale]{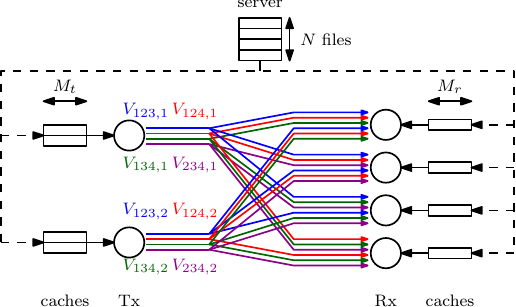}
}
\caption{The separation architecture applied to the setup in \figurename~\ref{fig:setup2x4} (i.e., Example~\ref{eg:2x4}) with multicast size $\kappa+1=3$.
The interface messages $V_{\mathcal{S}j}$ at the physical layer can be abstracted as orthogonal error-free multicast bit pipes at the network layer.
Thus at the physical layer (a) we focus on transmitting the $V_{\mathcal{S}j}$'s across the interference channel, while at the network layer (b) we perform the caching and delivery strategies, oblivious of the underlying physical channel, to deliver the requested files.}
\label{fig:separation}
\end{figure}

Our proposed separation architecture isolates the channel coding aspect of the problem from its content delivery aspect.
The former is handled at the physical layer, while the latter is handled at the network layer.
The two layers interface using a set $\mathscr{V}$ of multiple multicast messages,
\begin{equation}
\label{eq:v-set}
\mathscr{V} = \left\{
  V_{\mathcal{S}j} : j\in\{1,\ldots,K_t\}, \mathcal{S}\in\mathscr{S}
\right\},
\end{equation}
where $V_{\mathcal{S}j}$ denotes the message sent from transmitter $j$ to the subset $\mathcal{S}$ of receivers, and $\mathscr{S}\subseteq2^{\{1,\ldots,K_r\}}$ is some collection of subsets of receivers.
Notice that all transmitters have messages for the same subsets of receivers, a natural design choice due to the symmetry of the problem.
The physical layer processing transmits these messages across the interference network, while the network layer treats them as orthogonal error-free multicast bit pipes.
\figurename~\ref{fig:separation} illustrates this separation for the setting in Example~\ref{eg:2x4}.

In order to motivate our choice of $\mathscr{S}$ (and hence of $\mathscr{V}$), it will be useful to give a brief overview of the strategy used for the broadcast setup in \cite{maddah-ali2012}.
Suppose that the receiver memory is $M_r=\kappa N/K_r$, where $\kappa\in\{0,1,\ldots,K_r-1\}$ is an integer.
The idea is to place content in the receiver caches such that every subset of $\kappa$ of them shares an exclusive part of every file (each file is thus split into $\binom{K_r}{\kappa}$ equal parts).
During the delivery phase, linear combinations of these file parts are sent to every subset of $\kappa+1$ users such that each user can combine its received linear combination with the contents of its cache to decode one part of their requested file.
As a result, a total of
\begin{equation}
\label{eq:bc}
L^\text{BC}(N,K_r,M_r)\cdot F = \frac{K_r-\kappa}{\kappa+1} \cdot F
\end{equation}
bits are sent through the network (see \cite[Theorem 1]{maddah-ali2012}).

Notice that the broadcast strategy never really sends any \emph{broadcast} message on a logical level (except when $\kappa+1=K_r$).
Instead, it sends many \emph{multicast} messages, each intended for $\kappa+1$ users, which just happen to be ``overheard'' by the unintended receivers.
Inspired by this, we choose the messages in $\mathscr{V}$ to reflect the multicast structure in \cite{maddah-ali2012}.
Specifically, we choose to create one multicast message from each transmitter to every subset of receivers of size $\kappa+1$.
In other words,
\begin{equation}
\label{eq:s-set-multicast}
\mathscr{S} = \left\{
  \mathcal{S} \subseteq\{1,\ldots,K_r\} : |\mathcal{S}|=\kappa+1
\right\}.
\end{equation}
For example, \figurename~\ref{fig:separation} shows the separation architecture when $\kappa+1=3$.
While \eqref{eq:s-set-multicast} depicts the choice of $\mathscr{S}$ that we make most of the time, it is inefficient in a particular regime, namely when both the number of files and the receiver memory are small.
Since that regime is of only limited interest, we relegate its description to Appendix~\ref{app:small-n}.

Let $\tilde R_{\kappa+1}$ be the rate at which we transmit these messages at the physical layer, i.e., $V_{\mathcal{S}j}\in[2^{\tilde R_{\kappa+1}T}]$.
Further, let $\ell_{\kappa+1}$ be the size (normalized by file size) of whatever is sent through each multicast link at the network layer, i.e., $V_{\mathcal{S}j}\in[2^{\ell_{\kappa+1}F}]$.
Therefore, $\tilde R_{\kappa+1}T=\ell_{\kappa+1}F$.
Let us write $\tilde R_{\kappa+1}^\star$ and $\ell_{\kappa+1}^\star$ to denote the optimal $\tilde R_{\kappa+1}$ and $\ell_{\kappa+1}$, respectively, within their respective subproblems (these will be defined rigorously in the subsections below).
These quantities can be connected to the rate $R$ of the original caching problem.
Indeed, since $F=RT$, then we can achieve a rate $R$ equal to
\begin{equation}
\label{eq:achievability}
R = \frac{\tilde R_{\kappa+1}^\star}{\ell_{\kappa+1}^\star},
\end{equation}
when $M_r=\kappa N/K_r$, $\kappa\in\{0,1,\ldots,K_r-1\}$.%
\footnote{The nature of the separation architecture implies that $\kappa$ must always be an integer.
Regimes where it is not are handled using time and memory sharing between points where it is.
Furthermore, we exclude the case $\kappa=K_r$ (equivalently, $M_r=N$) for mathematical convenience, but we can in fact trivially achieve an infinite rate when $M_r=N$ by storing the complete content library in every user's cache.}

The separation architecture has thus created two subproblems of the original problem.
At the physical layer, we have a pure communication subproblem, where multicast messages $V_{\mathcal{S}j}$ must be transmitted reliably across an interference network.
At the network layer, we have a caching subproblem with noiseless orthogonal multicast links connecting transmitters to receivers.
In the two subsections below, we properly formulate each subproblem.
We give a strategy for each as well as the values of $\tilde R_{\kappa+1}$ and $\ell_{\kappa+1}$ that they achieve.

\subsection{Physical Layer}

At the physical layer, we consider only the communication problem of transmitting specific messages across the interference channel described in Section~\ref{sec:setup}, as illustrated in \figurename~\ref{fig:separation-phy}.
This is an interesting communication problem on its own, and we hence formulate it without all the caching details.
The message set that we consider is one where every transmitter $j$ has a message for every subset $\mathcal{S}$ of $\sigma$ receivers, where $\sigma\in\{1,\ldots,K_r\}$ is given.%
\footnote{In the context of the caching problem, $\sigma$ is chosen to be $\kappa+1$, as described earlier.}
We label such a message as $V_{\mathcal{S}j}$, and we note that there are a total of $K_t\binom{K_r}{\sigma}$ of them.
For instance, in the example shown in \figurename~\ref{fig:separation-phy}, message $V_{134,2}$ (used as a shorthand for $V_{\{1,3,4\},2}$) is sent by transmitter $2$ to receivers $1$, $3$, and $4$.
We call this problem the multiple multicast X-channel with multicast size $\sigma$, as it generalizes the (unicast) X-channel studied in \cite{cadambe2009} to multicast messages.
Note that, when $\sigma=1$, we recover the unicast X-channel.

We assume a symmetric setup, where all the messages have the same rate $\tilde R_\sigma$, i.e., $V_{\mathcal{S}j}\in[2^{\tilde R_\sigma T}]$.
A rate is called achievable if a strategy exists allowing all receivers to recover all their intended messages with vanishing error probability as the block length $T$ increases.
Our goal is to find the largest achievable rate $\tilde R_\sigma$ for a given $\SNR$, denoted by $\tilde R^\star_\sigma(\SNR)$, and in particular its DoF
\[
\tilde d^\star_\sigma(K_t,K_r)
\defeq \lim_{\SNR\to\infty} \frac{\tilde R^\star_\sigma(\SNR)}{\frac12\log\SNR}.
\]
One of the contributions of this paper is an exact characterization of $\tilde d^\star_\sigma$, and we next give an overview of how to achieve it.

For every receiver $i$, there is a set of $K_t\binom{K_r-1}{\sigma-1}$ desired messages $\{V_{\mathcal{S}j}:i\in \mathcal{S}\}$, and a set of $K_t\binom{K_r-1}{\sigma}$ interfering messages $\{V_{\mathcal{S}j}:i\notin \mathcal{S}\}$.
Using TDMA, all $K_t\binom{K_r}{\sigma}$ messages can be delivered to their receivers at a sum DoF of $1$, i.e., $\tilde d_\sigma = 1/K_t\binom{K_r}{\sigma}$.
However, by applying an interference alignment technique that generalizes the one used in \cite{cadambe2009}, we can, loosely speaking, collapse the $K_t\binom{K_r-1}{\sigma}$ interfering messages at every receiver into a subspace of dimension $\binom{K_r-1}{\sigma}$ (assuming for simplicity that each message forms a subspace of dimension one), while still allowing reliable recovery of all $K_t\binom{K_r-1}{\sigma-1}$ desired messages.
Thus an overall vector space of dimension $K_t\binom{K_r-1}{\sigma-1}+\binom{K_r-1}{\sigma}<K_t\binom{K_r}{\sigma}$ is used to deliver all $K_t\binom{K_r}{\sigma}$ messages.
This strategy achieves a DoF-optimal rate, as asserted by the following theorem.
\begin{theorem}
\label{thm:phy}
The DoF of the symmetric multiple multicast X-channel with multicast size $\sigma$ is given by
\[
\tilde d^\star_\sigma(K_t,K_r) = \frac{1}{K_t\binom{K_r-1}{\sigma-1}+\binom{K_r-1}{\sigma}}.
\]
\end{theorem}
The details of the interference alignment strategy are given in Section~\ref{sec:alignment}.
The proof of optimality is left for Appendix~\ref{app:phy-converse}, since it does not directly contribute to our main result in Theorem~\ref{thm:dof}.
It does however reinforce it by providing a complete solution to the physical-layer communication subproblem.

The DoF shown in Theorem~\ref{thm:phy} is a per-message DoF.
Since there are a total of $K_t\binom{K_r}{\sigma}$ messages, we obtain a sum DoF of
\begin{IEEEeqnarray*}{rCl}
K_t\binom{K_r}{\sigma} \cdot \tilde d_\sigma^\star(K_t,K_r)
&=& \frac{K_t\binom{K_r}{\sigma}}{K_t\binom{K_r-1}{\sigma-1}+\binom{K_r-1}{\sigma}}\\
&=& \frac{K_tK_r}{(K_t-1)\sigma+K_r}.
\IEEEyesnumber\label{eq:phy-sumdof}
\end{IEEEeqnarray*}
When $\sigma=1$, the sum DoF in \eqref{eq:phy-sumdof} is $K_tK_r/(K_t+K_r-1)$, thus recovering the unicast X-channel result from \cite{cadambe2009}.
When $\sigma=K_r$, the problem reduces to a broadcast channel with multiple sources, giving a sum DoF of~$1$.

\subsection{Network Layer}

The network layer setup is similar to the end-to-end setup, with the difference that the interference network is replaced by the multicast links $V_{\mathcal{S}j}$ from transmitters to receivers, as illustrated in \figurename~\ref{fig:separation-net}.
As mentioned previously, each link $V_{\mathcal{S}j}$ is shared by exactly $|\mathcal{S}|=\kappa+1$ users, where $\kappa=K_rM_r/N$ is an integer.
We again focus on a symmetric setup, where all links have the same size $\ell_{\kappa+1}$, called the \emph{link load}.
It will be easier in the discussion to use the \emph{sum network load} $L_{\kappa+1}$, i.e., the combined load of all $K_t\binom{K_r}{\kappa+1}$ links,
\begin{equation}
\label{eq:network-load}
L_{\kappa+1} = K_t\binom{K_r}{\kappa+1} \cdot \ell_{\kappa+1}.
\end{equation}

A sum network load $L$ is said to be achievable if, for every large enough file size $F$, a strategy exists allowing all users to recover their requested files with high probability while transmitting no more than $LF$ bits through the network.
Our goal is to find the smallest achievable network load for every $N$, $K_t$, $K_r$, $M_t$, and $M_r$, denoted by
\[
L_{\kappa+1}^\star(N,K_t,K_r,M_t,M_r),
\]
where $\kappa=K_rM_r/N$ is an integer.
Using a similar strategy to \cite{maddah-ali2012}, we achieve the following sum network load.
\begin{lemma}
\label{lemma:net}
In the network layer setup with a multicast size of $\kappa+1$, $\kappa\in\{0,1,\ldots,K_r-1\}$, a sum network load of
\[
L_{\kappa+1}^\star(N,K_t,K_r,M_t,M_r) \le \frac{K_r-\kappa}{\kappa+1}
\]
can be achieved when $M_r=\kappa N/K_r$.
\end{lemma}
\begin{IEEEproof}
We first divide every file $W_n$ into $K_t$ equal parts, $W_n=(W_n^1,\ldots,W_n^{K_t})$, and store the $j$-th part $W_n^j$ in the cache of transmitter $j$.
Note that, while we allow $M_t\ge N/K_t$ as per the regularity condition in \eqref{eq:mt-condition}, the above transmitter placement only stores exactly $N/K_t$ files at every transmitter irrespective of the value of $M_t$.
The different transmitters are then treated as independent sublibraries.
Indeed, the receiver placement splits each receiver cache into $K_t$ equal sections, and each section is dedicated to one sublibrary.
A placement phase identical to \cite{maddah-ali2012} is then performed for each sublibrary in its dedicated receiver memory.

During the delivery phase, user $i$'s request for a single file $W_{u_i}$ is converted into $K_t$ separate requests for the subfiles $(W_{u_i}^1,\ldots,W_{u_i}^{K_t})$, each from its corresponding sublibrary (transmitter).
For every subset $\mathcal{S}$ of $\kappa+1$ receivers, each transmitter $j$ then sends through the link $V_{\mathcal{S}j}$ exactly what would be sent to these receivers in the broadcast setup, had the other transmitters not existed.
This is possible since the $V_{\mathcal{S}j}$ links were chosen by design to match the multicast transmissions in the broadcast setup.
Each transmitter will thus send $(1/K_t)\cdot L^\text{BC}(N,K_r,M_r)$ files through the network (with $L^\text{BC}$ as defined in \eqref{eq:bc}), for a total network load of $(K_r-\kappa)/(\kappa+1)$.
\end{IEEEproof}

\subsection{Achievable End-to-End DoF}

From \eqref{eq:achievability} and using \eqref{eq:network-load}, we can achieve an end-to-end DoF of
\[
\DoF
\ge \frac{\tilde d_{\kappa+1}^\star(K_t,K_r)}{L^\star_{\kappa+1}(N,K_t,K_r,M_t,M_r)}
\cdot K_t\binom{K_r}{\kappa+1}.
\]
By combining Theorem~\ref{thm:phy} (with $\sigma=\kappa+1$) and Lemma~\ref{lemma:net},
\begin{IEEEeqnarray*}{rCl}
\frac{1}{\DoF}
&\le& \frac{L_{\kappa+1}^\star(N,K_t,K_r,M_t,M_r)}{\tilde d_{\kappa+1}^\star(K_t,K_r)}
\cdot \frac{1}{K_t\binom{K_r}{\kappa+1}}\\
&\le& \frac{K_r-\kappa}{\kappa+1}
\cdot \left[ K_t\binom{K_r-1}{\kappa} + \binom{K_r-1}{\kappa+1} \right]
\cdot \frac{1}{K_t\binom{K_r}{\kappa+1}}.
\end{IEEEeqnarray*}
By writing
\begin{IEEEeqnarray*}{rCl}
\IEEEeqnarraymulticol{3}{l}{K_t\binom{K_r-1}{\kappa} + \binom{K_r-1}{\kappa+1}}\\
\quad &=& (K_t-1)\binom{K_r-1}{\kappa} + \binom{K_r-1}{\kappa} + \binom{K_r-1}{\kappa+1}\\
&\overset{(a)}{=}& (K_t-1)\frac{\kappa+1}{K_r}\binom{K_r}{\kappa+1} + \binom{K_r}{\kappa+1}\\
&=& \frac{(K_t-1)(\kappa+1)+K_r}{K_r} \cdot \binom{K_r}{\kappa+1},
\end{IEEEeqnarray*}
where $(a)$ is due to Pascal's triangle, we conclude that
\begin{IEEEeqnarray*}{rCl}
\frac{1}{\DoF}
&\le& \frac{K_r-\kappa}{\kappa+1} \cdot \frac{(K_t-1)(\kappa+1)+K_r}{K_tK_r}\\
&=& \frac{K_t-1+\frac{K_r}{\kappa+1}}{K_t} \cdot \left( 1 - \frac{\kappa}{K_r} \right).
\IEEEyesnumber\label{eq:dof-multicast}
\end{IEEEeqnarray*}
This proves the achievability direction of Theorem~\ref{thm:dof} when $K_r/(\kappa+1)\le N$.
The case $K_r/(\kappa+1)>N$ is discussed in Appendix~\ref{app:small-n}.

%% file: input_files/alignment.tex
The multiple multicast X-channel problem (with multicast size $\sigma$) that emerges from our separation strategy is a generalization of the unicast ($\sigma=1$) X-channel studied in \cite{cadambe2009}.
We propose an interference alignment strategy that generalizes the one in \cite{cadambe2009}.
In this section, we give a high-level overview of the alignment strategy in order to focus on the intuition.
The rigorous explanation of the strategy is given in Appendix~\ref{app:alignment-details} as a proof of Lemma~\ref{lemma:phy-achievability}, which is presented at the end of this section.

Consider communicating across the interference network over $T$ time slots.
Every transmitter $j$ beamforms each message $V_{\mathcal{S}j}$ along some fixed vector of length $T$ and sends the sum of the vectors corresponding to all its messages as its codeword.
Each message thus occupies a subspace of dimension $1$ of the overall $T$-dimensional vector space.
The goal is to align at each receiver the interfering messages into the smallest possible subspace, so that a high rate is achieved for the desired messages.

When choosing which messages to align, we enforce the following three principles, which ensure maximal alignment without preventing decodability of the intended messages.
At every receiver $i$:
\begin{enumerate}
\item
Each desired message $V_{\mathcal{S}j}$ with $i\in \mathcal{S}$ must be in a subspace of dimension $1$, not aligned with any other subspace.
\item
Messages from the same transmitter must never be aligned.
\item
All messages intended for the same subset $\mathcal{S}$ of receivers with $i\notin \mathcal{S}$ must be aligned into one subspace of dimension $1$.
\end{enumerate}
Principle 1 ensures that receiver $i$ can decode all of its desired messages.
To understand principle 2, notice that messages from the same transmitter go through the same channels.
Therefore, if two messages from the same transmitter are aligned at one receiver, then they were also aligned during transmission, and are hence aligned at all other receivers, including their intended ones.
Thus principle 2 ensures decodability at other receivers.
As for principle 3, it provides the maximal alignment of the interfering messages without violating principle 2.
Indeed, each aligned subspace contains $K_t$ messages, one from each transmitter.
Any additional message that is aligned would share a transmitter with one of them.

For every receiver, there are $K_t\binom{K_r-1}{\sigma-1}$ desired messages.
By principle 1, each should take up one non-aligned subspace of dimension $1$, for a total of $K_t\binom{K_r-1}{\sigma-1}$ dimensions.
On the other hand, there are $K_t\binom{K_r-1}{\sigma}$ interfering messages.
By principle 3, every $K_t$ of them are aligned in one subspace of dimension~$1$, and hence all interfering messages fall in a subspace of dimension $\binom{K_r-1}{\sigma}$.
These subspaces can be made non-aligned by ensuring that the overall vector space has a dimension of
\[
T = K_t\binom{K_r-1}{\sigma-1}+\binom{K_r-1}{\sigma}.
\]
Since each message took up one dimension, we get a per-message DoF of
\[
\frac1T = \frac{1}{K_t\binom{K_r-1}{\sigma-1}+\binom{K_r-1}{\sigma}}.
\]
This is an improvement over TDMA, which achieves a DoF of $1/K_t\binom{K_r}{\sigma}$.

In most cases, we do not achieve the exact DoF shown in Theorem~\ref{thm:phy} using a finite number of channel realizations.
We instead achieve an arbitrarily close DoF by using an increasing number of channel realizations.
The exact achieved DoF is given in the following lemma.
\begin{lemma}
\label{lemma:phy-achievability}
Let $\Gamma=(K_r-\sigma)(K_t-1)$.
For any arbitrary $n\in\mathbb{N}$, we can achieve a DoF for message $V_{\mathcal{S}j}$ equal to
\[
\delta_j^{(n)}
= \frac{ (n+c_j)^\Gamma }
	{ \binom{K_r-1}{\sigma-1}\left[ (n+1)^\Gamma + (K_t-1)n^\Gamma \right]
		+ \binom{K_r-1}{\sigma}(n+1)^\Gamma
	},
\]
where $c_1=1$ and $c_2=\cdots=c_{K_t}=0$.
\end{lemma}
The proof of Lemma~\ref{lemma:phy-achievability} is given in Appendix~\ref{app:alignment-details}.

Note that Lemma~\ref{lemma:phy-achievability} achieves a slightly different DoF for $V_{\mathcal{S}j}$ depending on $j$, which might seem to contradict the symmetry in the problem setting.
However, for a large $n$, we have $(n+1)^\Gamma\approx n^\Gamma$, and hence
\[
\lim_{n\to\infty} \delta_j^{(n)}
=
\frac{1}{K_t\binom{K_r-1}{\sigma-1}+\binom{K_r-1}{\sigma}}
= \tilde d_\sigma^\star(K_t,K_r)
\]
for all $j$.
Thus the symmetric DoF $\tilde d_\sigma^\star(K_t,K_r)$ is achieved in the limit.

%% file: input_files/converse.tex
In this section, we give a high-level proof of the converse part of Theorem~\ref{thm:dof} by showing that the DoF achieved by the separation architecture in Section~\ref{sec:separation} is order-optimal.
We do this by computing cut-set-based information-theoretic upper bounds on the DoF (equivalently, they are lower bounds on the reciprocal $1/\DoF$).
These bounds are given in the following lemma, whose proof is placed at the end of this section in order not to distract from the intuition behind the converse arguments.
The rigorous converse proof is given in Appendix~\ref{app:converse}.

\begin{lemma}
\label{lemma:converse}
For any $N$, $K_t$, $K_r$, $M_t\in[0,N]$, and $M_r\in[0,N]$, the optimal DoF must satisfy
\[
\frac{1}{\DoF}
\ge \max_{s\in\{1,\ldots,\min\{K_r,N\}\}}
\frac{s\left(1-\frac{M_r}{\floor{N/s}}\right)}{\min\{s,K_t\}}.
\]
\end{lemma}

Lemma~\ref{lemma:converse} is next used to prove the converse part of Theorem~\ref{thm:dof}, i.e.,
\[
\DoF \le 13.5 \cdot d(N,K_t,K_r,M_t,M_r),
\]
where $d(\cdot)$ is defined in \eqref{eq:achievable}.
The procedure is similar to the one used in \cite{maddah-ali2012}: we consider three main regimes (Regimes 1, 2, and 3) of receiver memory $M_r$ and in each compare the expression $d(\cdot)$ with the outer bounds.
In addition, we consider a separate corner case (Regime 0) in which the largest possible number of distinct file requests (i.e., $\min\{K_r,N\}$) is small compared to the number of transmitters.
\begin{IEEEeqnarray*}{u?l}
Regime 0: & \min\{K_r,N\} \le 12.5 K_t;
	\IEEEyesnumber\label{eq:converse-regimes}
	\IEEEyessubnumber\\
Regime 1: & \min\{K_r,N\} > 12.5 K_t\\
	and & 0 \le M_r \le 1.1\max\left\{1,\frac{N}{K_r}\right\};
	\IEEEyessubnumber\\
Regime 2: & \min\{K_r,N\} > 12.5 K_t\\
	and & 1.1\max\left\{1,\frac{N}{K_r}\right\} < M_r \le 0.092 \frac{N}{K_t};
	\IEEEyessubnumber\\
Regime 3: & \min\{K_r,N\} > 12.5 K_t\\
	and & 0.092 \frac{N}{K_t} < M_r \le N.
	\IEEEyessubnumber
\end{IEEEeqnarray*}
Note that Regimes 1, 2, and 3 are unambiguous, since
\begin{IEEEeqnarray*}{rCl}
&& \min\{K_r,N\}>12.5K_t\\
&\implies&
0 < 1.1\max\left\{1,\frac{N}{K_r}\right\} < 0.092 \frac{N}{K_t} < N.
\IEEEyesnumber\label{eq:regimes-unambiguous}
\end{IEEEeqnarray*}

Since $M_r$ is the only variable that we will consistently vary, we will abuse notation for convenience and write $d(M_r)$ instead of $d(N,K_t,K_r,M_t,M_r)$ for all $M_r\in[0,N]$.
Our goal is thus to prove
\begin{equation}
\label{eq:converse}
\frac{1}{\DoF} \ge \frac{d^{-1}(M_r)}{13.5}.
\end{equation}
For ease of reference, we will rewrite the expression of $d^{-1}(M_r)$ here.
For $M_r=\kappa N/K_r$ where $\kappa\in\{0,1,\ldots,K_r\}$ is an integer,
\begin{equation}
\label{eq:reciprocal-expr}
d^{-1}(\kappa N/K_r)
= \frac{K_t - 1 + \min\left\{ \frac{K_r}{\kappa+1},N \right\}}{K_t}
	\cdot \left( 1 - \frac{\kappa}{K_r} \right),
\end{equation}
and $d^{-1}(M_r)$ is the lower convex envelope of these points for all $M_r\in[0,N]$.
Note that $d^{-1}(M_r)$ is non-increasing and convex in $M_r$.

\subsubsection*{Regimes 0 and 3}
Interestingly, Regimes 0 and 3 behave quite similarly to each other.
Indeed, notice that in both of them we have $K_t=\Omega\left(\min\left\{K_r/(\kappa+1),N\right\}\right)$.
Using \eqref{eq:reciprocal-expr}, this implies
\begin{IEEEeqnarray*}{rCl}
d^{-1}(M_r)
&\approx& \frac{K_t+\min\left\{\frac{K_r}{\kappa+1},N\right\}}{K_t} \cdot \left( 1 - \frac{M_r}{N} \right)\\
&=& \Theta\left( 1 - \frac{M_r}{N} \right).
\end{IEEEeqnarray*}
Conversely, we can apply Lemma~\ref{lemma:converse} with $s=1$ to get $1/\DoF\ge1-M_r/N$.
Thus in both regimes the local caching gain is the only significant contribution to the DoF.

\subsubsection*{Regime 1}
In Regime 1, the receiver memory is too small to have any significant effect.
Therefore, using $12.5K_t<\min\{K_r,N\}$, we can write \eqref{eq:reciprocal-expr} as
\begin{IEEEeqnarray*}{rCl}
d^{-1}(M_r)
&\approx& \frac{K_t+\min\left\{\frac{K_r}{\kappa+1},N\right\}}{K_t} \cdot 1\\
&\le& \left( \frac{1}{12.5} + 1 \right) \cdot \frac{\min\{K_r,N\}}{K_t}.
\end{IEEEeqnarray*}
Conversely, by using Lemma~\ref{lemma:converse} with $s\approx \min\{K_r,N\}$, we get
\[
\frac{1}{\DoF} \gtrsim \frac{s \cdot 1}{K_t} \approx \frac{\min\{K_r,N\}}{K_t}.
\]
Therefore, in this regime $\DoF\approx K_t/\min\{K_r,N\}$.
We can explain this in terms of the DoF gains in \eqref{eq:dof-gains}: when the receiver memory is very small, the only relevant gain is the interference alignment gain.

\subsubsection*{Regime 2}
In Regime 2, the receivers combined can store all of the content library.
As a result, the global caching gain kicks in.
We can upper-bound $d^{-1}(M_r)$ in \eqref{eq:reciprocal-expr} as follows:
\begin{IEEEeqnarray*}{rCl}
d^{-1}(M_r)
&\le& 1 + \frac{K_r}{K_t(K_rM_r/N+1)}
\le 1 + \frac{N}{K_tM_r}\\
&\le& 1.092 \frac{N}{K_tM_r},
\end{IEEEeqnarray*}
because $K_t<0.092N/M_r$ in Regime 2.
Conversely, let us apply Lemma~\ref{lemma:converse} using $s\approx N/2M_r$:
\[
\frac{1}{\DoF}
\gtrsim \frac{s-s^2M_r/N}{K_t}
\approx \frac{N}{4K_tM_r}.
\]
Therefore, $\DoF\approx K_tM_r/N$.
This behavior is similar to what one would expect in the broadcast setup in \cite{maddah-ali2012}, with the exception of the additional $K_t$ factor.

Since $d^{-1}(M_r)$ approximately matches the outer bounds in all four regimes and can also be achieved as in Section~\ref{sec:separation}, then it provides an approximate characterization of $1/\DoF$.
The above arguments are made rigorous in Appendix~\ref{app:converse}.

\begin{IEEEproof}[Proof of Lemma~\ref{lemma:converse}]
Consider $s\in\{1,\ldots,\min\{K_r,N\}\}$ users.
We shall look at $E=\floor{N/s}$ different request vectors, such that the combined number of files requested by all users after $E$ request instances is $\tilde N = sE=s\floor{N/s}$ files.
More specifically, we consider the request vectors $\mathbf{u}_1,\ldots,\mathbf{u}_E$ with
\[
\mathbf{u}_e = \bigl( \underbrace{(e-1)s + 1 , (e-1)s + 2 , \ldots , es}_s \,,\, \underbrace{1 \,,\, \ldots \,,\, 1}_{K_r-s} \bigr),
\]
for each $e=1,\ldots,E$.
Note that we only focus on the first $s$ users; the remaining $K_r-s$ users are not relevant to our argument.

When the request vector is $\mathbf{u}$, let $\mathbf{x}_j^{\mathbf{u}}$ and $\mathbf{y}_i^{\mathbf{u}}$ denote the inputs and outputs of the interference network for all transmitters $j$ and receivers $i$.
For notational convenience, we write $\mathbf{y}_{[s]}^\mathbf{u}=(\mathbf{y}_1^\mathbf{u},\ldots,\mathbf{y}_s^\mathbf{u})$ and use a similar notation for $\mathbf{x}_{[K_t]}^\mathbf{u}$.
Also, let $Q_i$ denote the contents of user $i$'s cache (recall that the cache contents are independent of $\mathbf{u}$).
By Fano's inequality,
\begin{equation}
\label{eq:fano}
H\left( W_1,\ldots,W_{\tilde N} \middle| Q_1,\ldots,Q_s, \mathbf{y}_{[s]}^{\mathbf{u}_1},\ldots,\mathbf{y}_{[s]}^{\mathbf{u}_E} \right) \le \epsilon T,
\end{equation}
since the $s$ users should be able to each decode their $\floor{N/s}$ requested files using their caches and channel outputs.
Then,
\begin{IEEEeqnarray*}{rCl}
\tilde NRT
&=& H\left( W_1,\ldots,W_{\tilde N} \right)\\
&=& I\left( W_1,\ldots,W_{\tilde N} ; Q_1,\ldots,Q_s, \mathbf{y}_{[s]}^{\mathbf{u}_1},\ldots,\mathbf{y}_{[s]}^{\mathbf{u}_E} \right)\\
&& {} + H\left( W_1,\ldots,W_{\tilde N} \middle| Q_1,\ldots,Q_s, \mathbf{y}_{[s]}^{\mathbf{u}_1},\ldots,\mathbf{y}_{[s]}^{\mathbf{u}_E} \right)\\
&\overset{(a)}{\le}& I\left( W_1,\ldots,W_{\tilde N} ; Q_1,\ldots,Q_s, \mathbf{y}_{[s]}^{\mathbf{u}_1},\ldots,\mathbf{y}_{[s]}^{\mathbf{u}_E} \right)
+ \epsilon T\\
&=& I\left( W_1,\ldots,W_{\tilde N} ; \mathbf{y}_{[s]}^{\mathbf{u}_1},\ldots,\mathbf{y}_{[s]}^{\mathbf{u}_E} \right)\\
&& {} + I\left( W_1,\ldots,W_{\tilde N} ; Q_1,\ldots,Q_s \middle| \mathbf{y}_{[s]}^{\mathbf{u}_1},\ldots,\mathbf{y}_{[s]}^{\mathbf{u}_E} \right)\\
&& {} + \epsilon T\\
&\le& I\left( W_1,\ldots,W_{\tilde N} ; \mathbf{y}_{[s]}^{\mathbf{u}_1},\ldots,\mathbf{y}_{[s]}^{\mathbf{u}_E} \right)\\
&& {} + H\left( Q_1,\ldots,Q_s \right)
+ \epsilon T\\
&\overset{(b)}{\le}& I\left( \mathbf{x}_{[K_t]}^{\mathbf{u}_1},\ldots,\mathbf{x}_{[K_t]}^{\mathbf{u}_E} ;
  \mathbf{y}_{[s]}^{\mathbf{u}_1},\ldots,\mathbf{y}_{[s]}^{\mathbf{u}_E} \right)\\
&& {} + sM_rRT
+ \epsilon T\\
&\overset{(c)}{\le}& E\cdot\max_{e\in\{1,\ldots,E\}}
I\left( \mathbf{x}_1^{\mathbf{u}_e},\ldots,\mathbf{x}_{K_t}^{\mathbf{u}_e} ;
  \mathbf{y}_1^{\mathbf{u}_e},\ldots,\mathbf{y}_{s}^{\mathbf{u}_e} \right)\\
&& {} + sM_rRT
+ \epsilon T\\
&\overset{(d)}{\le}& E\cdot T \left(
  \min\left\{K_t,s\right\} \cdot \frac12\log\SNR + o(\log\SNR)
\right)\\
&& {} + sM_rRT + \epsilon T\\
&=& \floor{N/s}T \left(
  \min\left\{K_t,s\right\} \cdot \frac12\log\SNR + o(\log\SNR)
\right)\\
&& {}+ sM_rRT + \epsilon T,
\end{IEEEeqnarray*}
where $(a)$ is due to inequality \eqref{eq:fano}, $(b)$ uses the data processing inequality, $(c)$ follows from the independence of the channel outputs when conditioned on all channel inputs, and $(d)$ is the capacity bound of the $K_t\times s$ MIMO channel over $T$ time blocks.

Since $\tilde N=s\floor{N/s}$, and by taking $T\to\infty$ and $\epsilon\to0$, we obtain
\[
R\left( 1 - \frac{M_r}{\floor{N/s}} \right)
\le \frac1s \cdot \min\{K_t,s\} \cdot \frac12\log\SNR + o(\log\SNR).
\]
The DoF thus obeys
\[
\DoF \le \frac{\min\{K_t,s\}}{s\left(1-\frac{M_r}{\floor{N/s}}\right)}.
\]
Since $s$ was arbitrary, the above is true for any $s\in\{1,\ldots,\min\{K_r,N\}\}$, and thus the lemma is proved.
\end{IEEEproof}

%% file: input_files/2x2.tex
In this paper, we have determined the approximate DoF of the general cache-aided interference network.
To do so, we have proposed a separation-based strategy that uses interference alignment to create \emph{non-interacting} multicast bit pipes from transmitters to receivers, and we have shown that this strategy achieves a DoF that is within a constant multiplicative factor from the optimum.
However, this achieved DoF is only approximately optimal.
In fact, many improvements can be made, such as using transmit zero-forcing as has been discussed in previous work \cite{maddah-ali2015interference,naderializadeh2016,xu2016}.

\begin{figure}
\centering
\includegraphics[scale=\myfigsscale]{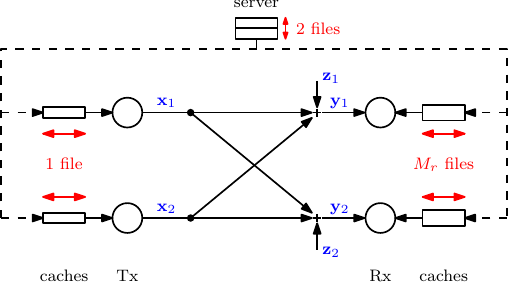}
\caption{The $2\times2$ cache-aided interference channel with $2$ files.
The transmitter caches can hold exactly one file each, and the receiver caches $M_r\in[0,2]$ files each.
The $\mathbf{z}_i$'s are iid additive Gaussian unit-variance noise.}
\label{fig:2x2-setup}
\end{figure}

In this section, we explore a different approach, which lies within the context of interference alignment described in Section~\ref{sec:alignment}: rather than ignoring the interference subspace, which contains the aligned messages, we attempt to extract some information from it.
Thus every receiver gains additional information in the form of an alignment of the bit pipes available at other receivers: the bit pipes would thus \emph{interact}.
We study this approach in a specific setup: the $2\times2$ interference channel with a content library of two files, shown in \figurename~\ref{fig:2x2-setup}.

\begin{figure}
\centering
\includegraphics[scale=\myfigsscale]{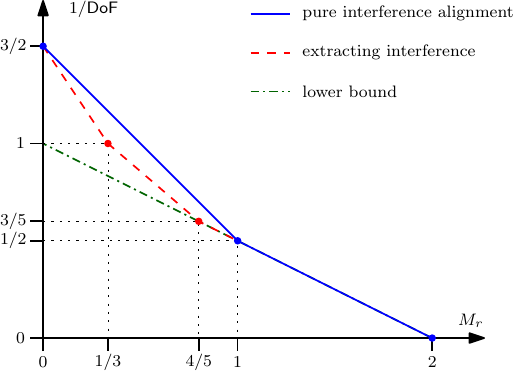}
\caption{Inverse DoF achieved by the scheme from Section~\ref{sec:alignment} (solid line), and the improved inverse DoF achieved by extracting more information from the aligned interference (dashed line).
The dash-dotted line shows the information-theoretic lower bounds from Lemma~\ref{lemma:converse}.}
\label{fig:2x2-dof}
\end{figure}

For this $2\times2$ setup, by Theorem~\ref{thm:dof} the main strategy described in this paper achieves
\[
\frac{1}{\DoF} \le \max\left\{
\frac32-M_r,
1 - \frac12 M_r
\right\},
\]
for $M_r\in[0,2]$, as shown by the solid line in \figurename~\ref{fig:2x2-dof}.
However, the same figure shows an improved inverse DoF, depicted by the dashed line, which is achieved using the interference-extracting scheme discussed in this section.
A factor-$7/6$ improvement is obtained over the main strategy.
This result is summarized in the following theorem.
\begin{theorem}
\label{thm:2x2}
The following inverse DoF can be achieved for the $2\times2$ cache-aided interference network with $N=2$ files and transmitter memory $M_t=1$:
\[
\frac{1}{\DoF} \le \max\left\{
\frac32 - \frac32 M_r,
\frac97 - \frac67 M_r,
1 - \frac12 M_r
\right\},
\]
for all values of $M_r\in[0,2]$.
\end{theorem}

It should be noted that the general converse stated in Lemma~\ref{lemma:converse} can be applied here and results in
\[
\frac{1}{\DoF} \ge 1 - \frac12 M_r,
\]
which implies that our strategy is exactly optimal for $M_r\ge4/5$, as illustrated by the dash-dotted line in \figurename~\ref{fig:2x2-dof}.

\begin{figure}
\centering
\subfloat[Physical-layer view.]{%
	\label{fig:2x2-phy}
	\includegraphics[scale=\myfigsscale]{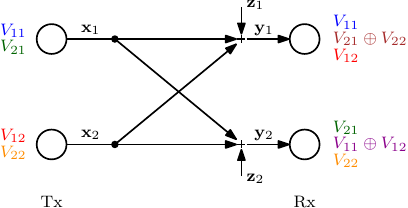}
}
\\
\subfloat[Network-layer view.]{%
	\label{fig:2x2-net}
	\includegraphics[scale=\myfigsscale]{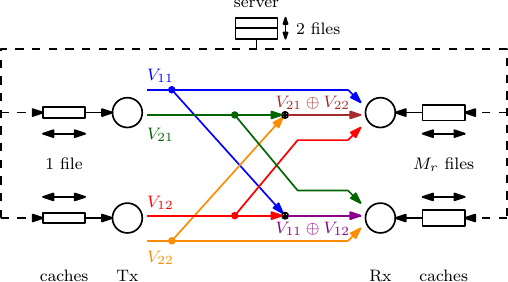}
}
\caption{Separation architecture with interference extraction in the $2\times2$ case with $2$ files.
The (unicast) X-channel message set is used, but every receiver decodes, in addition to its intended messages, the sum of the messages intended for the other reciever.}
\label{fig:2x2-separation}
\end{figure}

We will next give a high-level overview of the interference-extraction strategy.
The proof of Theorem~\ref{thm:2x2}, including the details of the strategy, are given in Appendix~\ref{app:2x2}.
Consider what happens when the main strategy is used in this $2\times2$ setup with $M_r=0$.
The strategy creates one unicast message from every transmitter to every receiver, and transmits them using interference alignment.
Each receiver thus gets the two messages intended for it, plus an alignment of the two messages intended for the other receiver.
In the main strategy, this aligned interference is simply discarded.
However, we can design the scheme in a way that this alignment is a simple sum of the two interfering messages.
Each receiver can then decode, in addition to its intended messages, the sum of the interfering messages, without suffering any decrease in the sum DoF of the communicated messages.
We hence obtain a new separation architecture, illustrated in \figurename~\ref{fig:2x2-separation}, that we use for all $M_r$.

The scheme we propose in this section is very specific to the $2\times2$ interference network with two files in the content library.
An interesting direction for future work would be to extend this interference-extraction strategy to more general settings.

%% file: input_files/discussion.tex
In this paper, we have presented the approximate degrees of freedom of cache-aided interference networks, with caches at both the transmitters and the receivers.
While an exact characterization of the DoF is certainly desirable, finding it is a more difficult problem since the exact rate-memory trade-off is unknown even for the error-free broadcast case.

The DoF can be approximately achieved using the separation architecture described in Section~\ref{sec:separation}, which decouples the physical-layer transmission scheme from the network-layer coded caching scheme.
While this strategy is approximately optimal, some improvements can still be made, albeit with no more than a constant-factor gain.
We explored one such improvement in Section~\ref{sec:2x2} where the aligned subspaces that result from the physical-layer interference alignment scheme are extracted and used as additional bit pipes at the receivers.

\begin{figure}
\centering
\includegraphics[width=.48\textwidth]{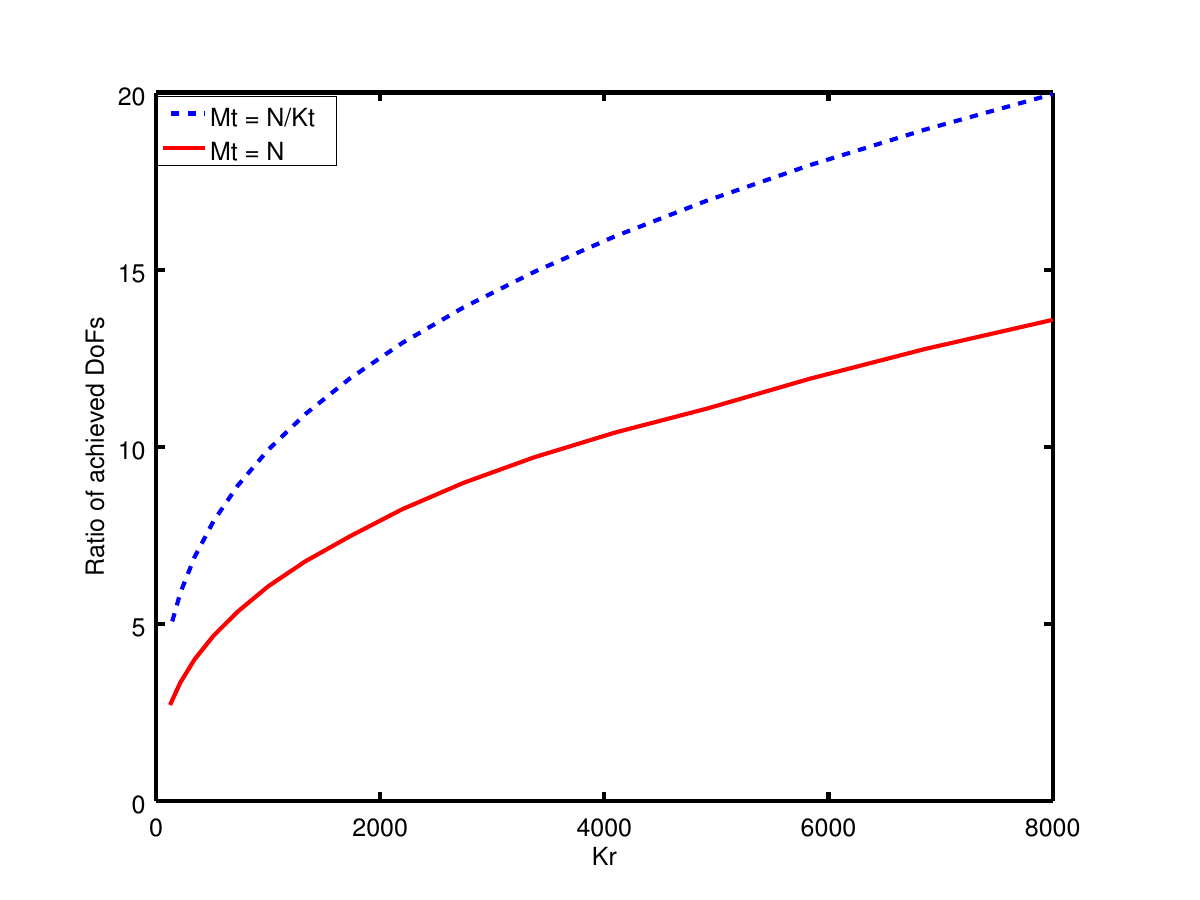}
\caption{Largest ratio of the DoF achieved by our proposed scheme to the DoF achieved by the one-shot linear scheme proposed in \cite{naderializadeh2016}.
In this figure, the number of receivers $K_r$ is scaled, while $K_t = K_r^{1/3}$.
The plot shows the maximum ratio between the DoFs over all possible receiver memory values $M_r\in[0,N]$.
The comparison is made for two values of the transmitter memory, $M_t=N/K_t$ and $M_t=N$.
Notice that the gap increases arbitrarily with $K_r$.}
\label{fig:linear-scheme-compare}
\end{figure}

In the literature, a similar setting to the one in this paper was recently studied in \cite{naderializadeh2016}.
However, since \cite{naderializadeh2016} focuses on one-shot linear schemes, the interference alignment gain is not achieved.
This significantly reduces the achieved degrees of freedom, especially in the lower memory regime when the number of receivers is large.
In particular, if $M_r=N/\sqrt{K_r}$ and $K_t\le\sqrt{K_r}$, then we can show that the DoF achieved by our scheme is larger than the one-shot linear scheme by a multiplicative factor of at least
\[
\frac{K_tK_r}{(K_t+\sqrt{K_r})^2} \ge \frac{K_t}{4},
\]
which can be arbitrarily large.
A tighter comparison is numerically illustrated in \figurename~\ref{fig:linear-scheme-compare} for $K_t=K_r^{1/3}$ and $M_t$ taking the values $N/K_t$ and $N$.

Possible extensions to the problem include further improvements to the scheme, such as by using transmit zero-forcing or by placing coded content in the caches; a derivation of tighter outer bounds; and an exploration of the regime where the total transmitter memory is less than the size of the content library, i.e., $N-M_r\le K_tM_t<N$.
Since the initial posting of our paper on arXiv in June 2016, several follow-up works have extended the results in a few of these directions \cite{xu2016-arxiv,roig2017}.
Another interesting question is to find the (exactly) optimal strategy when the problem imposes a restriction of uncoded cache placement, in a similar manner to \cite{wan-uncoded2016,yu-uncoded2017} for the broadcast case.

%% file: input_files/small-n.tex
Recall that the separation architecture creates a set of messages $\mathscr{V}$ as an interface between the physical and network layers,
\[
\mathscr{V} = \left\{ V_{\mathcal{S}j} : j\in\{1,\ldots,K_t\}, \mathcal{S}\in\mathscr{S} \right\}
\]
for some $\mathscr{S}\subseteq2^{\{1,\ldots,K_r\}}$, as seen in \eqref{eq:v-set}.
In this paper, we have so far focused on the choice of messages described by $\mathscr{S}$ in \eqref{eq:s-set-multicast}, in which every transmitter has a message for every subset of exactly $\kappa+1$ receivers, where $\kappa=K_rM_r/N$ is an integer.
While this is order-optimal in most cases, it is insufficient when both the receiver memory and the number of files is small.

To illustrate, consider the case with only a single file in the content library ($N=1$) and without receiver caches ($M_r=0$).
Furthermore, assume that there is just one transmitter ($K_t=1$) but many receivers ($K_r$ is large).
Seeing as there is only one file, all receivers will request that same file, and hence the obvious strategy is for the transmitter to broadcast the file to all receivers, thus achieving a DoF of $1$.
However, under the separation architecture described by \eqref{eq:s-set-multicast}, we create one message from the transmitter for every individual receiver, and then send that file separately as $K_r$ different messages.
This is clearly inefficient since the same file is being sent $K_r$ times, thus achieving a much worse DoF of $1/K_r$.

The reason the usual separation architecture is inefficient in this example is that it inherently assumes that all users request different files in the worst case.
This is true when there are more files than users.
However, if there are so few files that many users will inevitably request the same file, then the previous assumption fails.
In this appendix, we handle that case by providing a different separation interface.
We exclusively work with the case $M_r=0$ and compute an achievable DoF for it.
Specifically, we show that
\begin{equation}
\label{eq:dof-mreq0}
\frac{1}{\DoF} \le \frac{K_t+\min\{K_r,N\}-1}{K_t}.
\end{equation}
Since we also know that $1/\DoF$, which is convex in $M_r$, is zero when $M_r=N$, then we can achieve any linear combination of the two reciprocal DoFs between these two points, using time- and memory-sharing.
Specifically, we achieve
\begin{equation}
\label{eq:small-n-ach}
\frac{1}{\DoF} \le \frac{K_t+\min\{K_r,N\}-1}{K_t} \cdot \left( 1 - \frac{M_r}{N} \right).
\end{equation}

The expression of the reciprocal DoF in \eqref{eq:small-n-ach} can be decomposed into two gains, in a similar way as in \eqref{eq:dof-gains}.
Since the strategy that achieves \eqref{eq:small-n-ach} is relevant when $N<K_r$, we can write the two gains as
\[
N\DoF
\approx
\underbrace{ \frac{K_tN}{K_t+N-1} }_{g^\mathrm{IA}}
\cdot \underbrace{ \frac{1}{1 - \frac{M_r}{N}} }_{g^\mathrm{LC}}.
\]
Note that $N\DoF$ is the sum DoF here since the total number of requested files is $N<K_r$ in the worst case.

The most striking difference with \eqref{eq:dof-gains} is that there is no global caching gain.
Indeed, the strategy makes no use of any coding or multicasting opportunities, as we will see below.
On the other hand, the local caching gain is present and is the same as before.
The interference alignment gain is slightly different: it is the interference alignment gain of a $K_t\times N$ unicast X-channel, not $K_t\times K_r$.
The reason for this is that, when $N<K_r$, then the total number of \emph{distinct} requested files is $N$ in the worst case.
The strategy thus only needs to account for $N$ distinct demands, and uses methods from the compound X-channel \cite{maddah-ali-compound,gou2011} to serve them.

We proceed with the strategy for $M_r=0$ that achieves \eqref{eq:dof-mreq0}.
Since $M_r=0$, we cannot store anything in the receiver caches.
In the transmitter caches, we place the same content as previously described, i.e., every file $W_n$ is split into $K_t$ parts, and transmitter $j$ stores the $j$-th part of $W_n$, called $W_n^j$, for every $n$.
In the delivery phase, we partition the set of users into subsets such that all the users in the same subset request the same file.
Specifically, let $\mathbf{u}$ denote the request vector, and let $\mathcal{U}_n=\{i:u_i=n\}$ be the set of users requesting file $W_n$.
Our goal is to create a multicast message from every transmitter to all users that are requesting the same file.
In other words, we set
\[
\mathscr{S} = \left\{ \mathcal{U}_n : n\in\{1,\ldots,N\} \text{ s.t. } \mathcal{U}_n\not=\emptyset \right\}
\]
Note that $\mathscr{S}$ is a partition of the entire set of users.
We denote its size by $\tilde N=|\mathscr{S}|$, which is equivalent to the total number of distinct requested files.
Our separation interface $\mathscr{V}$ is thus a set of messages from every transmitter to $\tilde N$ \emph{non-overlapping} subsets of receivers,
\[
\mathscr{V} = \left\{ V_{\mathcal{U}_nj} : \mathcal{U}_n\not=\emptyset \text{ and } j=1,\ldots,K_t \right\}.
\]
We focus on transmitting these messages across the interference channel at the physical layer.
At the network layer, we use these messages as error-free bit pipes to deliver the requested files to the users at the network layer.

\subsection{Physical Layer}

At the physical layer, the problem is equivalent to the compound X-channel problem, described in \cite{maddah-ali-compound,gou2011}.
In the $K_t\times K_r$ compound X-channel, every transmitter has a message for every receiver.
However, the channel of every receiver $i$ can be one of some \emph{finite} number $J_i$ of states, and transmission has to account for all possible states.
The optimal sum DoF in this problem is $K_tK_r/(K_t+K_r-1)$, i.e., $1/(K_t+K_r-1)$ per message, as stated in \cite[Theorem~4]{maddah-ali-compound}.

If the receiver is able to decode its messages regardless of which of the $J_i$ realizations the channel has taken, then this is equivalent to replacing the single receiver with $J_i$ channel realizations by $J_i$ different receivers with each a single possible channel realization, such that all $J_i$ receivers want the same messages.
This is exactly the problem statement we have at the physical layer.
Our problem is therefore equivalent to a $K_t\times\tilde N$ compound X-channel with $|\mathcal{U}_n|$ channel realizations for every receiver $n$.
Therefore, if $\tilde R_{\tilde N}$ denotes the rate of each message, and $\tilde d_{\tilde N}$ its DoF, then \cite[Theorem~4]{maddah-ali-compound} implies that the optimal DoF is
\begin{equation}
\label{eq:compound-dof}
\tilde d_{\tilde N}(K_t,K_r) = \frac{1}{K_t+\tilde N-1}.
\end{equation}

\subsection{Network Layer}

Let the link load $\ell_{\tilde N}$ denote the size of each $V_{\mathcal{U}_nj}$ in units of files.
The strategy at the network layer is straightforward.
For every subset $\mathcal{U}_n$ of users, each transmitter $j$ sends the part of the file that they requested through $V_{\mathcal{U}_nj}$.
Mathematically, we set
\[
V_{\mathcal{U}_nj} = W_n^j,
\]
for all $j=1,\ldots,K_t$ and $n$ such that $\mathcal{U}_n\not=\emptyset$.
This allows every user to decode its requested file.
Since every multicast link $V_{\mathcal{U}_nj}$ carries one file part $W_n^j$, then the link load is
\begin{equation}
\label{eq:partition-load}
\ell_{\tilde N}(N,K_t,K_r,M_t,0) = \frac{1}{K_t}.
\end{equation}

\subsection{Achievable End-to-End DoF}

Note that the same $V_{\mathcal{U}_nj}$ has a size of $\tilde R_{\tilde N}T$ at the physical layer and $\ell_{\tilde N}F$ at the network layer.
Since $F=RT$, we get $R=\tilde R_{\tilde N}/\ell_{\tilde N}$, and by combining that with \eqref{eq:compound-dof} and \eqref{eq:partition-load}, we achieve a DoF of
\[
\frac{\tilde d_{\tilde N}(K_t,K_r)}{\ell_{\tilde N}(N,K_t,K_r,M_t,0)} = \frac{K_t}{K_t+\tilde N-1}.
\]
In the worst case, the largest number of distinct files are requested, i.e., $\tilde N=\min\{K_r,N\}$.
Therefore,
\[
\DoF \ge \frac{K_t}{K_t+\min\{K_r,N\}-1},
\]
when $M_r=0$.

Since $1/\DoF$ is convex in $M_r$, and $1/\DoF=0$ when $M_r=N$, then, for all intermediate values of $M_r$, we can achieve
\begin{equation}
\label{eq:dof-partition}
\frac{1}{\DoF} \le \frac{K_t+\min\{K_r,N\}-1}{K_t} \cdot \left( 1 - \frac{M_r}{N} \right).
\end{equation}

For any $M_r\in[0,N]$, we can choose whichever of the two separation interfaces yields the higher DoF.
Therefore, combining \eqref{eq:dof-partition} with \eqref{eq:dof-multicast} yields
\begin{IEEEeqnarray*}{rCll}
\frac{1}{\DoF}
&\le& \min\Biggl\{&
	\frac{K_t+\min\{K_r,N\}-1}{K_t} \cdot \left( 1 - \frac{\kappa}{K_r} \right),\\
&&& \frac{K_t-1+\frac{K_r}{\kappa+1}}{K_t} \cdot \left( 1 - \frac{\kappa}{K_r} \right)
\Biggr\}\\
&=& \IEEEeqnarraymulticol{2}{l}{
	  \frac{K_t+\min\left\{ \frac{K_r}{\kappa+1},N \right\} - 1}{K_t}
	  \cdot \left( 1 - \frac{\kappa}{K_r} \right),}
\end{IEEEeqnarray*}
when $M_r=\kappa N/K_r$ with $\kappa$ being an integer, and the lower convex envelope of these points for all $M_r\in[0,N]$.
This concludes the achievability proof of Theorem~\ref{thm:dof}.

%% file: input_files/alignment-details.tex
Let $\Gamma$ be defined as in the statement of the lemma, and let $n\in\mathbb{N}$ be arbitrary.
Define $T_n$ as
\begin{IEEEeqnarray*}{rCl}
T_n
&=& \binom{K_r-1}{\sigma-1}\left[ (n+1)^\Gamma + (K_t-1)n^\Gamma \right]\\
&& {} + \binom{K_r-1}{\sigma}(n+1)^\Gamma.
\end{IEEEeqnarray*}
We will show that a DoF of $\delta_j^{(n)}$ can be achieved for message $V_{\mathcal{S}j}$ over a block length of $T_n$.
We first describe how to (maximally) align the interference at each receiver, and then show that the receiver's desired messages are still decodable.
The proof will rely on two lemmas from \cite{cadambe2009}: the alignment part will use \cite[Lemma~2]{cadambe2009}, while the decodability part will rely on \cite[Lemma~1]{cadambe2009}.
For ease of reference, we have rephrased the two lemmas in Appendix~\ref{app:cadambe-lemmas} as Lemmas~\ref{lemma:alignment} and~\ref{lemma:decodability}, respectively.

\paragraph*{Alignment}
Describe each message $V_{\mathcal{S}j}$ as a column vector of $(n+c_j)^\Gamma$ symbols $\mathbf{v}_{\mathcal{S}j}=[v_{\mathcal{S}j}^m]_{m=1}^{(n+c_j)^\Gamma}$, where $c_j$ is as defined in the statement of the lemma.
Each symbol is beamformed along a length-$T_n$ vector $\mathbf{a}_{\mathcal{S}j}^m$, so that transmitter $j$ sends the codeword
\[
\mathbf{x}_j = \sum_{\mathcal{S}:|\mathcal{S}|=\sigma} \sum_{m=1}^{(n+c_j)^\Gamma} v_{\mathcal{S}j}^m\mathbf{a}_{\mathcal{S}j}^m,
\]
over the block length $T_n$.
We can alternatively combine all the $\mathbf{a}_{\mathcal{S}j}^m$ vectors into one matrix
$\mathbf{A}_{\mathcal{S}j}=[\mathbf{a}_{\mathcal{S}j}^1,\ldots,\mathbf{a}_{\mathcal{S}j}^{(n+c_j)^\Gamma}]$,
and write
\[
\mathbf{x}_j = \sum_{\mathcal{S}} \mathbf{A}_{\mathcal{S}j}\mathbf{v}_{\mathcal{S}j}.
\]

Receiver $i$ then observes
\begin{equation}
\label{eq:phy-ach-yi}
\mathbf{y}_i
= \sum_{j=1}^{K_t} \mathbf{H}_{ij}\mathbf{x}_j + \mathbf{z}_i
= \sum_{j=1}^{K_t} \mathbf{H}_{ij}\sum_{\mathcal{S}} \mathbf{A}_{\mathcal{S}j} \mathbf{v}_{\mathcal{S}j} + \mathbf{z}_i.
\end{equation}
Recall that $\mathbf{z}_i$ is the iid additive Gaussian unit-variance noise, and $\mathbf{H}_{ij}$ is a $T_n\times T_n$ diagonal matrix representing the independent continuously-distributed channel coefficients over block length $T_n$, as defined in Section~\ref{sec:setup}.
In other words, the $\tau$-th diagonal element of $\mathbf{H}_{ij}$ is $h_{ij}(\tau)$.
Moreover, the dimensions of $\mathbf{A}_{\mathcal{S}j}$ are $T_n\times(n+c_j)^\Gamma$, and the length of $\mathbf{v}_{\mathcal{S}j}$ is $(n+c_j)^\Gamma$.

In the expression for $\mathbf{y}_i$ in \eqref{eq:phy-ach-yi}, it will be convenient to separate the messages intended for $i$ from the interfering messages,
\begin{IEEEeqnarray*}{rCl}
\mathbf{y}_i
&=& \sum_{\mathcal{S}:i\in \mathcal{S}} \sum_{j=1}^{K_t} \mathbf{H}_{ij}\mathbf{A}_{\mathcal{S}j}\mathbf{v}_{\mathcal{S}j}\\
&& {} + \sum_{\mathcal{S}:i\notin \mathcal{S}} \left[
	\mathbf{H}_{i1}\mathbf{A}_{\mathcal{S}1}\mathbf{v}_{\mathcal{S}1}
	+ \sum_{j=2}^{K_t} \mathbf{H}_{ij}\mathbf{A}_{\mathcal{S}j}\mathbf{v}_{\mathcal{S}j}
\right]
+ \mathbf{z}_i.
\end{IEEEeqnarray*}
Our goal is to collapse each term inside the second sum (i.e., for each $\mathcal{S}$ such that $i\notin \mathcal{S}$) into a single subspace, namely the subspace spanned by $\mathbf{H}_{i1}\mathbf{A}_{\mathcal{S}1}$.%
\footnote{This is why we choose $\mathbf{v}_{\mathcal{S}1}$ to be a longer vector than $\mathbf{v}_{\mathcal{S}j},j\ge2$: this choice makes $\mathbf{H}_{i1}\mathbf{A}_{\mathcal{S}1}$ the larger subspace, which allows us to align the other subspaces with it using Lemma~\ref{lemma:alignment}.}
This should be done for all $i\in\{1,\ldots,K_r\}$.
Specifically, we want to choose the $\mathbf{A}_{\mathcal{S}j}$'s such that they satisfy the following conditions almost surely:
\begin{IEEEeqnarray*}{r'l}
\mathbf{H}_{ij}\mathbf{A}_{\mathcal{S}j} \prec \mathbf{H}_{i1}\mathbf{A}_{\mathcal{S}1},
& \forall i=1,\ldots,K_r,\\
& \forall j=2,\ldots,K_t,\\
& \forall \mathcal{S} \text{ s.t. } i\notin \mathcal{S},
\end{IEEEeqnarray*}
where $\mathbf{P}\prec\mathbf{Q}$ denotes that the vector space spanned by the columns of $\mathbf{P}$ is a subspace of the one spanned by the columns of $\mathbf{Q}$.

First, we set $\mathbf{A}_{\mathcal{S}2} = \cdots = \mathbf{A}_{\mathcal{S}K_t}$ for all subsets $\mathcal{S}$.
Thus we have reduced the problem to finding matrices $\mathbf{A}_{\mathcal{S}1}$ and $\mathbf{A}_{\mathcal{S}2}$ for all subsets $\mathcal{S}$ such that, almost surely,
\begin{equation}
\label{eq:phy-ach-subspace}
\mathbf{H}_{i1}^{-1}\mathbf{H}_{ij}\mathbf{A}_{\mathcal{S}2} \prec \mathbf{A}_{\mathcal{S}1},
\qquad \forall i\notin \mathcal{S},
\quad  \forall j=2,\ldots,K_t.
\end{equation}
Note that $\mathbf{H}_{i1}^{-1}$ exists almost surely since each diagonal element of $\mathbf{H}_{i1}$ follows a continuous distribution and is thus non-zero with probability one.

For every $\mathcal{S}$, the matrices $\mathbf{A}_{\mathcal{S}1}$ and $\mathbf{A}_{\mathcal{S}2}$ are constrained by a total of $(K_r-\sigma)(K_t-1)=\Gamma$ subspace relations.
We hence have $\Gamma$ relations $\mathbf{G}_{g}\mathbf{A}_{\mathcal{S}2}\prec\mathbf{A}_{\mathcal{S}1}$, $g=1,\ldots,\Gamma$, where $\mathbf{G}_g$ are $T_n\times T_n$ diagonal matrices.
We can write all the diagonal elements of these matrices as forming the set
\[
\mathcal{G} = \left\{
\frac{h_{ij}(\tau)}{h_{i1}(\tau)} : i\notin \mathcal{S}, j\in\{2,\ldots,K_t\}, \tau\in\{1,\ldots,T_n\}
\right\}.
\]
Importantly, each element of $\mathcal{G}$ follows a continuous distribution when conditioned on all the others.
In other words,
\[
\left.
\frac{h_{ij}(\tau)}{h_{i1}(\tau)}
\middle|
\left\{ \frac{h_{i'j'}(\tau')}{h_{i'1}(\tau')} : (i',j',\tau')\not=(i,j,\tau) \right\}
\right.
\]
obeys a continuous distribution.
Furthermore, the dimensions of $\mathbf{A}_{\mathcal{S}2}$ are $T_n\times n^\Gamma$, and the dimensions of $\mathbf{A}_{\mathcal{S}1}$ are $T_n\times(n+1)^\Gamma$, with $T_n>(n+1)^\Gamma$.

Let $P$ be some continuous probability distribution with a bounded support.
For every $\mathcal{S}\subseteq\{1,\ldots,K_r\}$ such that $|\mathcal{S}|=\sigma$, we generate a $T_n\times1$ column vector $\mathbf{b}_\mathcal{S}=(b_\mathcal{S}(1),\ldots,b_\mathcal{S}(T_n))^\top$, such that all the entries of all $\binom{K_r}{\sigma}$ vectors $\{\mathbf{b}_\mathcal{S}\}_\mathcal{S}$ are chosen iid from $P$.
We can now invoke Lemma~\ref{lemma:alignment} in Appendix~\ref{app:cadambe-lemmas} to construct with probability one, for each $\mathcal{S}$ and using $\mathbf{b}_\mathcal{S}$, full-rank matrices $\mathbf{A}_{\mathcal{S}1}$ and $\mathbf{A}_{\mathcal{S}2}$ that satisfy the subspace relations in \eqref{eq:phy-ach-subspace} almost surely.
Furthermore, the entries of the $\tau$-th rows of both $\mathbf{A}_{\mathcal{S}1}$ and $\mathbf{A}_{\mathcal{S}2}$ are each a multi-variate monomial in the entries of the $\tau$-th rows of $\mathbf{b}_\mathcal{S}$ and $\mathbf{G}_g$, $g=1,\ldots,\Gamma$, i.e.,
\begin{equation}
\label{eq:phy-ach-monomials}
b_\mathcal{S}(\tau)
\quad\text{and}\quad
\frac{h_{ij}(\tau)}{h_{i1}(\tau)}, i\notin \mathcal{S}, j=2,\ldots,K_t.
\end{equation}
Note that the monomial entries of $\mathbf{A}_{\mathcal{S}1}$ are distinct; the same goes for the monomial entries of $\mathbf{A}_{\mathcal{S}2}$.

We have thus ensured alignment of the interfering messages.
In the remainder of the proof we show that the desired messages are still almost surely decodable at every receiver.

\paragraph*{Decodability}
Recall that the total dimension of the vector space, i.e., the block length, is
\[
T_n
= \binom{K_r-1}{\sigma-1}\left[ (n+1)^\Gamma + (K_t-1)n^\Gamma \right]
+ \binom{K_r-1}{\sigma}.
\]
Let us fix a receiver $k$.
For this receiver, we have:
\begin{itemize}
\item
$\binom{K_r-1}{\sigma-1}$ subspaces $\mathbf{H}_{k1}\mathbf{A}_{\mathcal{S}1}$, $k\in \mathcal{S}$, of dimension $(n+1)^\Gamma$ each, carrying the length-$(n+1)^\Gamma$ vectors $\mathbf{v}_{\mathcal{S}1}$ that must be decoded by receiver $k$;
\item
$(K_t-1)\binom{K_r-1}{\sigma-1}$ subspaces $\mathbf{H}_{kj}\mathbf{A}_{\mathcal{S}2}$, $k\in \mathcal{S}$ and $j=2,\ldots,K_t$, of dimension $n^\Gamma$ each, carrying the length-$n^\Gamma$ vectors $\mathbf{v}_{\mathcal{S}2}$ that must also be decoded by receiver $k$;
\item
$\binom{K_r-1}{\sigma}$ subspaces corresponding to $\mathbf{H}_{k1}\mathbf{A}_{\mathcal{S}1}$, $k\notin \mathcal{S}$, of dimension $(n+1)^\Gamma$ each, collectively carrying all the interference at receiver $k$.
\end{itemize}
Our goal is to show that the above subspaces are non-aligned for every receiver $k$, which implies that the desired messages are decodable with high probability for a large enough SNR.

Define matrices $\mathbf{D}_{k}$ and $\mathbf{I}_k$ representing the subspaces carrying the desired messages and the interference, respectively, by horizontally concatenating the subspaces:
\begin{IEEEeqnarray*}{rCl}
\mathbf{D}_k &=& \begin{bmatrix}
	\mathbf{H}_{k1}\mathbf{A}_{\mathcal{S}1}
	& \mathbf{H}_{k2}\mathbf{A}_{\mathcal{S}2}
	& \cdots
	& \mathbf{H}_{k{K_t}}\mathbf{A}_{\mathcal{S}2}
\end{bmatrix}_{\mathcal{S}:k\in \mathcal{S}};\quad
\IEEEyesnumber\label{eq:phy-ach-dkik}\IEEEyessubnumber\\
\mathbf{I}_k &=&
\begin{bmatrix}
	\mathbf{H}_{k1}\mathbf{A}_{\mathcal{S}1}
\end{bmatrix}_{\mathcal{S}:k\notin \mathcal{S}}.
\IEEEyessubnumber
\end{IEEEeqnarray*}
Therefore, decodability at receiver $k$ is ensured if the $T_n\times T_n$ matrix
\[
\mathbf{\Psi}_k = \begin{bmatrix} \mathbf{D}_k & \mathbf{I}_k \end{bmatrix}
\]
is full rank almost surely.
We prove that this is true with the help of Lemma~\ref{lemma:decodability} in Appendix~\ref{app:cadambe-lemmas}.
To apply Lemma~\ref{lemma:decodability}, we need to show that the following two conditions hold.
\begin{enumerate}
\item
Two distinct rows of $\mathbf{\Psi}_k$ consist of monomials in disjoint sets of variables.
In other words, the variables involved in the monomials of a specific row are exclusive to that row.

\item
Within each row, each entry is a \emph{unique} product of powers of the variables associated with that row.
\end{enumerate}

To show that the first condition holds, consider the $\tau$-th row of $\mathbf{\Psi}_k$.
This row consists of monomial terms in the variables $\{b_{\mathcal{S}}(\tau)\}_{\mathcal{S}}$ and $\{h_{ij}(\tau)\}_{i,j}$.
This is true because the $\tau$-th row of any submatrix $\mathbf{H}_{kj}\mathbf{A}_{\mathcal{S}j}$ of $\mathbf{\Psi}_k$ is equal to $h_{kj}(\tau)$ multiplied by the $\tau$-th row of $\mathbf{A}_{\mathcal{S}j}$, whose entries are monomials in the variables listed in \eqref{eq:phy-ach-monomials}.
Therefore, the variables involved in a row of $\mathbf{\Psi}_k$ are exclusive to that row.

Before we prove that the second condition holds, we emphasize two remarks regarding the monomials that constitute the entries of the $\tau$-th row of $\mathbf{\Psi}_k$.
\begin{remark}
\label{rmk:same-row}
All the entries in the $\tau$-th row of submatrix $\mathbf{H}_{kj}\mathbf{A}_{\mathcal{S}j}$ are distinct monomials from one another.
This is true because the $\tau$-th row of $\mathbf{H}_{kj}\mathbf{A}_{\mathcal{S}j}$ is equal to the $\tau$-th row of $\mathbf{A}_{\mathcal{S}j}$, whose entries are distinct monomials by construction (using Lemma~\ref{lemma:decodability}), multiplied by $h_{kj}(\tau)$.
\end{remark}
\begin{remark}
\label{rmk:different-rows}
It follows from \eqref{eq:phy-ach-monomials} that the entries of the $\tau$-th row of submatrix $\mathbf{H}_{kj}\mathbf{A}_{\mathcal{S}j}$ are monomials in which only the variables in a set $\mathcal{B}_{\mathcal{S}j}(\tau)$ appear (with non-zero exponent), where $\mathcal{B}_{\mathcal{S}j}(\tau)$ obeys:
\begin{IEEEeqnarray*}{lCl}
\IEEEeqnarraymulticol{3}{l}{
b_{\mathcal{S}}(\tau) \in \mathcal{B}_{\mathcal{S}j}(\tau)
\quad\text{and}\quad
b_{\mathcal{S}'}(\tau) \notin \mathcal{B}_{\mathcal{S}j}(\tau),\
\forall \mathcal{S}'\not=\mathcal{S};}
\IEEEyesnumber\label{eq:phy-ach-monomials-full}
\IEEEyessubnumber\label{eq:phy-ach-monomials-full-S}\\
k\in\mathcal{S}
\implies
\mathcal{B}_{\mathcal{S}j}(\tau)
&=& \left\{
  b_{\mathcal{S}}(\tau),
  h_{kj}(\tau)
\right\}\\
&& {} \cup \left\{
  h_{i1}(\tau),\ldots,h_{iK_t}(\tau) : i\notin\mathcal{S}
\right\}.\\
\IEEEyessubnumber\label{eq:phy-ach-monomials-full-k}
\end{IEEEeqnarray*}
Note that when $k\notin\mathcal{S}$, we cannot be sure if $\mathcal{B}_{\mathcal{S}j}(\tau)$ contains $h_{kj}(\tau)$ because the latter is present in the monomials of both $\mathbf{H}_{kj}$ and $\mathbf{A}_{Sj}$, and can hence be canceled out in their product.
\end{remark}

Remark~\ref{rmk:same-row} states that monomials in the same submatrix are distinct.
Therefore, all that remains is to show the same for monomials in the $\tau$-th rows of different submatrices.
However, by Remark~\ref{rmk:different-rows} the same variables appear with non-zero exponent in \emph{all} entries in the $\tau$-th row of any submatrix (albeit with different powers).
It is therefore sufficient to prove that the $\tau$-th rows of two different submatrices $\mathbf{H}_{kj}\mathbf{A}_{\mathcal{S}j}$ and $\mathbf{H}_{kj'}\mathbf{A}_{\mathcal{S}'j'}$ are functions of different sets of variables.
Specifically, we show that there is a variable that appears with non-zero exponent in all the entries of the $\tau$-th row of one submatrix but in none of the entries of the $\tau$-th row of the other.
We will prove below that this claim is true using Remark~\ref{rmk:different-rows}, with the aid of Table~\ref{tbl:phy-ach-monomials} for visualization.

\begin{table*}
\centering
\caption{Exponents of variables in monomials of the $\tau$-th row of $\mathbf{\Psi}_k$ for an arbitrary receiver $k$.
The subsets $\mathcal{S}_1$, $\mathcal{S}_1'$, $\mathcal{S}_2$, and $\mathcal{S}_2'$ are arbitrary such that $k\in \mathcal{S}_1\cap \mathcal{S}_1'$ and $k\notin \mathcal{S}_2\cup \mathcal{S}_2'$.
The transmitters $j$ and $j'$ are also arbitrary.
A cell will contain a check mark (\checkmark) if the variable in the corresponding row appears with non-zero exponent in the monomials of the $\tau$-th row of the submatrix in the corresponding column.
The cell will be empty if the variable does not appear in those monomials.
It will contain a question mark (?) if the variable may or may not appear.
Not all variables and submatrices are shown; only a representative few are used.
Finally, recall that $\mathbf{A}_{\mathcal{S}j}=\mathbf{A}_{\mathcal{S}2}$ for $j\ge2$.}
\label{tbl:phy-ach-monomials}
\begin{tabular}{|r||c|c||c|c|c|c|}
\hline
& \multicolumn{2}{c||}{$\mathbf{I}_k$} & \multicolumn{4}{c|}{$\mathbf{D}_k$}\\
& $\mathbf{H}_{k1}\mathbf{A}_{\mathcal{S}_21}$ & $\mathbf{H}_{k1}\mathbf{A}_{\mathcal{S}_2'1}$
& $\mathbf{H}_{kj}\mathbf{A}_{\mathcal{S}_1j}$ & $\mathbf{H}_{kj}\mathbf{A}_{\mathcal{S}_1'j}$
& $\mathbf{H}_{kj'}\mathbf{A}_{\mathcal{S}_1j'}$ & $\mathbf{H}_{kj'}\mathbf{A}_{\mathcal{S}_1'j'}$\\
\hline
\hline
$b_{\mathcal{S}_1}(\tau)$    & & & \checkmark & & \checkmark &\\
\hline
$b_{\mathcal{S}_1'}(\tau)$ & & & & \checkmark & & \checkmark\\
\hline
$b_{\mathcal{S}_2}(\tau)$    & \checkmark & & & & &\\
\hline
$b_{\mathcal{S}_2'}(\tau)$ & & \checkmark & & & &\\
\hline
\hline
$h_{kj}(\tau)$  & ? & ? & \checkmark & \checkmark & &\\
\hline
$h_{kj'}(\tau)$ & ? & ? & & & \checkmark & \checkmark\\
\hline
\end{tabular}
\end{table*}

For convenience, define $\mathbf{r}_{\mathcal{S}j}^\top$ to be the $\tau$-th row of matrix $\mathbf{H}_{kj}\mathbf{A}_{\mathcal{S}j}$, and similarly define $\mathbf{r}_{\mathcal{S}'j'}^\top$ to be the $\tau$-th row of matrix $\mathbf{H}_{kj'}\mathbf{A}_{\mathcal{S}'j'}$.
To show that the entries of these rows are monomial functions of distinct variables, we isolate two cases: case $\mathcal{S}\not=\mathcal{S}'$ and case $\mathcal{S}=\mathcal{S}'$, $j\not=j'$.
\begin{enumerate}
\item
Suppose $\mathcal{S}\not=\mathcal{S}'$.
Then, by \eqref{eq:phy-ach-monomials-full-S}, $\mathbf{r}_{\mathcal{S}j}^\top$ is a function of $b_{\mathcal{S}}(\tau)$ but not $b_{\mathcal{S}'}(\tau)$ while the opposite is true of $\mathbf{r}_{\mathcal{S}'j'}^\top$.
\item
Suppose now that $\mathcal{S}=\mathcal{S}'$ but $j\not=j'$.
Crucially, two such matrices are relevant at receiver $k$ only if $k\in\mathcal{S}$, as evidenced by \eqref{eq:phy-ach-dkik}.
Therefore, by \eqref{eq:phy-ach-monomials-full-k}, row $\mathbf{r}_{\mathcal{S}j}^\top$ is a function of $h_{kj}(\tau)$ but not $h_{kj'}(\tau)$, and the reverse is true of $\mathbf{r}_{\mathcal{S}'j'}^\top$.
\end{enumerate}
Combining the above two points, it follows that the entries of the two rows are monomials in a different set of variables.
We can conclude that the entries in the $\tau$-th row of $\mathbf{\Psi}_k$ are distinct monomials, and specifically that the matrix $\mathbf{\Psi}_k$ is of the form seen in the statement of Lemma~\ref{lemma:decodability}.
Therefore, by Lemma~\ref{lemma:decodability}, the matrix $\mathbf{\Psi}_k$ is full rank almost surely, and thus all receivers are able to decode their desired messages almost surely.

\paragraph*{In conclusion}
We were able to transmit all the messages $V_{\mathcal{S}j}$, represented by length-$(n+c_j)^\Gamma$ vectors $\mathbf{v}_{\mathcal{S}j}$, over a block length of $T_n$.
Hence, the degrees of freedom achieved for each $V_{\mathcal{S}j}$ is
\begin{IEEEeqnarray*}{rCl}
\IEEEeqnarraymulticol{3}{l}{\frac{(n+c_j)^\Gamma}{T_n}}\\
\quad &=& \frac{(n+c_j)^\Gamma}
	{\binom{K_r-1}{\sigma-1}\left[(n+1)^\Gamma+(K_t-1)n^\Gamma\right]
	 +\binom{K_r-1}{\sigma}(n+1)^\Gamma}\\
&=& \delta_j^{(n)},
\end{IEEEeqnarray*}
which concludes the proof of Lemma~\ref{lemma:phy-achievability}.
\hfill\IEEEQED

%% file: input_files/converse-detailed.tex
A high-level overview of the converse proof of Theorem~\ref{thm:dof} was given in Section~\ref{sec:converse}.
In this appendix, we will give the rigorous proof.
In particular, we will prove \eqref{eq:converse}, i.e., 
\[
\frac{1}{\DoF} \ge \frac{d^{-1}(M_r)}{13.5}
\]
by analyzing the four regimes described in \eqref{eq:converse-regimes}.

\subsection*{Regime 0: $\min\{K_r,N\}\le12.5K_t$}
In this regime, the number of transmitters is at least of the order of the total number of different requested files.
As described in Section~\ref{sec:scaling-kt}, this implies that $1/\DoF\approx1-\frac{M_r}{N}$.
More precisely, by convexity of $d^{-1}(M_r)$ we have
\begin{IEEEeqnarray*}{rCl}
d^{-1}(M_r)
&\le& d^{-1}(0) - \frac{d^{-1}(0)-d^{-1}(N)}{N-0} (M_r-0)\\
&=& d^{-1}(0) \left( 1 - \frac{M_r}{N} \right)
\end{IEEEeqnarray*}
for all $M_r\in[0,N]$, where we have used that $d^{-1}(N)=0$.
Moreover, we have
\[
d^{-1}(0) = \frac{K_t-1+\min\{K_r,N\}}{K_t} \le \frac{K_t + 12.5K_t}{K_t} = 13.5,
\]
which implies
\begin{equation}
\label{eq:reg0-ach}
d^{-1}(M_r) \le 13.5\left( 1 - \frac{M_r}{N} \right).
\end{equation}

We now invoke Lemma~\ref{lemma:converse} with $s=1$, yielding
\begin{equation}
\label{eq:reg0-gap}
\frac{1}{\DoF} \ge 1 - \frac{M_r}{N}
\overset{(a)}{\ge} \frac{d^{-1}(M_r)}{13.5} \qquad\text{in Regime 0,}
\end{equation}
where $(a)$ follows from \eqref{eq:reg0-ach}.

In all the following regimes, $\min\{K_r,N\}>12.5K_t\ge12.5$.

\subsection*{Regime 1: $0\le M_r\le1.1\max\{1,N/K_r\}$}

Since $d^{-1}(M_r)$ is non-increasing in $M_r$, we can upper-bound it by
\begin{IEEEeqnarray*}{rCl}
d^{-1}(M_r)
&\le& d^{-1}(0)\\
&=& \frac{K_t-1+\min\{K_r,N\}}{K_t}\\
&\le& \frac{ \left( \frac{1}{12.5}+1 \right)\min\{K_r,N\} }{K_t}\\
&=& \frac{13.5}{12.5}\cdot\frac{\min\{K_r,N\}}{K_t}.
\IEEEyesnumber\label{eq:reg1-ach}
\end{IEEEeqnarray*}

Let us now use Lemma~\ref{lemma:converse} with $s=\floor{0.275\min\{K_r,N\}}\in\{1,\ldots,\min\{K_r,N\}\}$.
Then, using $\floor{N/s}\ge N/s-1$,
\begin{IEEEeqnarray*}{rCl}
& \IEEEeqnarraymulticol{2}{l}{\frac{1}{\DoF}}\\
&\ge& \frac{s}{\min\{s,K_t\}} \left( 1 - \frac{s}{1-s/N}\cdot\frac{M_r}{N} \right)\\
&\ge& \frac{1}{K_t} \left( s - \frac{s^2}{1-s/N}\cdot\frac{1.1\max\{1,N/K_r\}}{N} \right)\\
&=& \frac{1}{K_t} \biggl(
	\floor{0.275\min\{K_r,N\}}\\
&&{} - \frac{\floor{0.275\min\{K_r,N\}}^2}{1-\floor{0.275\min\{K_r,N\}}/N}
	\cdot \frac{1.1\max\{1,N/K_r\}}{N}
\biggr)\\
&\ge& \frac{1}{K_t} \biggl(
	0.275\min\{K_r,N\} - 1\\
&&{} - \frac{(0.275\min\{K_r,N\})^2}{1-0.275}
	\cdot \frac{1.1\max\{1,N/K_r\}}{N}
\biggr)\\
&=& \frac{1}{K_t} \biggl(
	0.275\min\{K_r,N\} - 1\\
&&{} - \frac{(0.275\min\{K_r,N\})^2}{0.725}
	\cdot \frac{1.1}{\min\{K_r,N\}}
\biggr)\\
&=& \frac{\min\{K_r,N\}}{K_t} \biggl(
	0.275 - \frac{1}{\min\{K_r,N\}} - \frac{(0.275)^2\cdot1.1}{0.725}
\biggr)\\
&\ge& \frac{\min\{K_r,N\}}{K_t} \biggl(
	0.275 - \frac{1}{12.5} - \frac{(0.275)^2\cdot1.1}{0.725}
\biggr)\\
&\ge& \frac{1}{12.5}\cdot\frac{\min\{K_r,N\}}{K_t}\\
&\overset{(a)}{\ge}& \frac{d^{-1}(M_r)}{13.5} \qquad\text{in Regime 1,}
\IEEEyesnumber\label{eq:reg1-gap}
\end{IEEEeqnarray*}
where $(a)$ uses \eqref{eq:reg1-ach}.

\subsection*{Regime 2: $1.1\max\{1,N/K_r\}<M_r\le0.092 N/K_t$}
Let $\tilde M_r$ be the largest integer multiple of $N/K_r$ that is no greater than $M_r$, and define $\tilde \kappa=K_r\tilde M_r/N$.
Note that $\tilde \kappa$ is an integer.
Hence,
\[
0 \le M_r - N/K_r < \tilde M_r \le M_r.
\]
Since $d^{-1}(M_r)$ is non-increasing in $M_r$, we have:
\begin{IEEEeqnarray*}{rCl}
d^{-1}(M_r)
&\le& d^{-1}(\tilde M_r)\\
&=& \frac{K_t-1+\min\{\frac{K_r}{\tilde\kappa+1},N\}}{K_t} \cdot \left( 1 - \frac{\tilde\kappa}{K_r} \right)\\
&\le& \frac{K_t+\frac{K_r}{\tilde\kappa+1}}{K_t}\\
&=& 1 + \frac{K_r}{K_t(\tilde\kappa+1)} \IEEEyesnumber\label{eq:reg2-ach-step}\\
&=& 1 + \frac{K_r}{K_t(K_r\tilde M_r/N+1)}\\
&\overset{(a)}{\le}& 1 + \frac{K_r}{K_t\cdot K_rM_r/N}\\
&=& \frac{N}{K_tM_r} \left( \frac{K_tM_r}{N} + 1 \right)\\
&\overset{(b)}{\le}& 1.092 \frac{N}{K_tM_r}, \IEEEyesnumber\label{eq:reg2-ach}
\end{IEEEeqnarray*}
where $(a)$ uses $\tilde M_r>M_r-N/K_r$ and $(b)$ follows from $M_r\le0.092N/K_t$.

We now invoke Lemma~\ref{lemma:converse} with $s=\floor{0.3N/M_r}\in\{1,\ldots,\min\{K_r,N\}\}$.
Once again, we write:
\begin{IEEEeqnarray*}{rCl}
\frac{1}{\DoF}
&\ge& \frac{1}{\min\{s,K_t\}}\left( s - \frac{s^2}{1-s/N}\cdot\frac{M_r}{N} \right)\\
&\ge& \frac{1}{K_t} \left(
	0.3\frac{N}{M_r} - 1
	- \frac{ (0.3N/M_r)^2 }{1 - 0.3/M_r}
	\cdot \frac{M_r}{N}
\right)\\
&\overset{(a)}{\ge}& \frac{1}{K_t} \left(
	0.3\frac{N}{M_r} - 1
	- \frac{ 0.3^2N/M_r }{1 - 0.3/1.1}
\right)\\
&=& \frac{N}{K_tM_r} \left( 0.3 - \frac{M_r}{N} - \frac{0.3^2}{1-0.3/1.1} \right)\\
&\overset{(b)}{\ge}& \frac{N}{K_tM_r} \left( 0.3 - 0.092 - \frac{0.3^2}{1-0.3/1.1} \right)\\
&\ge& \frac{1.092}{13.5} \cdot \frac{N}{K_tM_r}\\
&\overset{(c)}{\ge}& \frac{d^{-1}(M_r)}{13.5} \qquad\text{in Regime 2,}
\IEEEyesnumber\label{eq:reg2-gap}
\end{IEEEeqnarray*}
where $(a)$ is due to $M_r\ge1.1\max\{1,N/K_r\}\ge1.1$, $(b)$ follows from $M<0.092N/K_t\le0.092N$, and $(c)$ uses \eqref{eq:reg2-ach}.

\subsection*{Regime 3: $0.092N/K_t<M_r\le N$}
By the convexity of $d^{-1}(M_r)$, we have for all $M_r\in(0.092N/K_t,N]$,
\begin{IEEEeqnarray*}{rCl}
\IEEEeqnarraymulticol{3}{l}{d^{-1}(M_r)}\\
\quad&\le& d^{-1}(0.092N/K_t)\\
&&{} - \frac{d^{-1}(0.092N/K_t)-d^{-1}(N)}{N-0.092N/K_t} \cdot \left( M_r - 0.092\frac{N}{K_t} \right)\\
&\overset{(a)}{=}& d^{-1}(0.092N/K_t) \left( 1 - \frac{M_r-0.092N/K_t}{N-0.092N/K_t} \right)\\
&=& d^{-1}(0.092N/K_t) \left( \frac{N-M_r}{N-0.092N/K_t} \right)\\
&=& \frac{d^{-1}(0.092N/K_t)}{1-0.092/K_t} \left( 1 - \frac{M_r}{N} \right)\\
&\le& \frac{d^{-1}(0.092N/K_t)}{0.908} \left( 1 - \frac{M_r}{N} \right),
\IEEEyesnumber\label{eq:reg3-ach-step1}
\end{IEEEeqnarray*}
where $(a)$ uses that $d^{-1}(N)=0$.

Let $\tilde M_r$ be the largest integer multiple of $N/K_r$ that is no greater than $0.092N/K_t$, and define $\tilde\kappa=K_r\tilde M_r/N$.
Note that $\tilde\kappa$ is an integer.
Then,
\[
0 \overset{(a)}{\le} 0.092N/K_t - N/K_r < \tilde M_r \le 0.092N/K_t,
\]
where $(a)$ follows from \eqref{eq:regimes-unambiguous}.
This implies that $\tilde\kappa+1\ge0.092K_r/K_t$.
Since $d^{-1}(M_r)$ is non-increasing,
\begin{IEEEeqnarray*}{rCl}
d^{-1}(0.092N/K_t)
&\le& d^{-1}(\tilde M_r)\\
&\overset{(a)}{\le}& 1 + \frac{K_r}{K_t(\tilde\kappa+1)}\\
&\le& 1 + \frac{K_r}{K_t\cdot 0.092K_r/K_t}\\
&=& \frac{1.092}{0.092}, \IEEEyesnumber\label{eq:reg3-ach-step2}
\end{IEEEeqnarray*}
where $(a)$ follows the same steps that led to \eqref{eq:reg2-ach-step}.
Combining \eqref{eq:reg3-ach-step2} with \eqref{eq:reg3-ach-step1},
\begin{equation}
\label{eq:reg3-ach}
d^{-1}(M_r)
\le \frac{1.092}{0.092\cdot0.908} \left( 1 - \frac{M_r}{N} \right)
\le 13.1 \left( 1 - \frac{M_r}{N} \right).
\end{equation}

By applying Lemma~\ref{lemma:converse} with $s=1$ again, we obtain
\begin{equation}
\label{eq:reg3-gap}
\frac{1}{\DoF}
\ge 1 - \frac{M_r}{N}
\overset{(a)}{\ge} \frac{d^{-1}(M_r)}{13.5} \qquad\text{in Regime 3,}
\end{equation}
where $(a)$ follows from \eqref{eq:reg3-ach}.

\subsection*{Synthesis}
The inequalities in \eqref{eq:reg0-gap}, \eqref{eq:reg1-gap}, \eqref{eq:reg2-gap}, and \eqref{eq:reg3-gap} cover all possible regimes.
Therefore, they together give $1/\DoF \ge d^{-1}(M_r)/13.5$, or equivalently
\[
\DoF \le 13.5 \cdot d\left( N,K_t,K_r,M_t,M_r \right),
\]
for all $N$, $K_t$, $K_r$, $M_t\in[0,N]$, and $M_r\in[0,N]$.
This concludes the converse proof of Theorem~\ref{thm:dof}.
\hfill\IEEEQED

%% file: input_files/phy-converse.tex
In Section~\ref{sec:separation} we have described the separation architecture and the communication problem that emerges from it.
We call the communication problem the multiple multicast X-channel problem.
We state its DoF $\tilde d_\sigma^\star(K_t,K_r)$ in Theorem~\ref{thm:phy} and show that it is achievable by interference alignment in Section~\ref{sec:alignment}.
In this appendix, we prove its optimality by deriving matching information-theoretic outer bounds.
Specifically, we want to prove
\begin{equation}
\label{eq:dof-converse}
\tilde d_\sigma^\star(K_t,K_r)
\le \frac{1}{K_t\binom{K_r-1}{\sigma-1}+\binom{K_r-1}{\sigma}},
\end{equation}
for all $K_t$ and $K_r$.

Consider the following subset of messages:
\begin{equation}
\label{eq:phy-v}
\mathcal{V} = \left\{ V_{\mathcal{S}j} : 1\in \mathcal{S} \text{ or } j=1 \right\}.
\end{equation}
It will be convenient to split $\mathcal{V}$ into two disjoint parts,
\begin{IEEEeqnarray*}{rCl}
\mathcal{V}^r &=& \left\{ V_{\mathcal{S}j} : 1\in \mathcal{S}, j\in\{1,\ldots,K_t\} \right\},\\
\mathcal{V}^t &=& \left\{ V_{\mathcal{S}j} : 1\notin \mathcal{S}, j=1 \right\}.
\end{IEEEeqnarray*}
In what follows, we will only focus on $\mathcal{V}$.
All other messages, collectively denoted by
\[
\overline{\mathcal{V}} = \left\{ V_{\mathcal{S}j} : 1\notin \mathcal{S} \text{ and } j\not=1 \right\},
\]
are made available to everyone by a genie.
Furthermore, we lower the noise at receiver one by a fixed (non-vanishing) amount.
Specifically, we replace $\mathbf{y}_1$ by
\[
\tilde{\mathbf{y}}_1 = \sum_{j=1}^{K_t} \mathbf{H}_{ij}\mathbf{x}_j + \tilde{\mathbf{z}}_1,
\]
where $\tilde z_1(\tau)$ are independent zero-mean Gaussian variables with variance
\begin{equation}
\label{eq:phy-noise-variance}
\var\left(\tilde z_1(\tau)\right) = \min_{i=1,\ldots,K_r} \left( \frac{h_{11}(\tau)}{h_{i1}(\tau)} \right)^2.
\end{equation}
Note that $\var(\tilde z_1(\tau))\le1$ since we can set $i=1$ in \eqref{eq:phy-noise-variance}.
Hence all the above changes can only improve capacity.

Consider all the receivers other than receiver one.
Let a genie also give all of these receivers the subset $\mathcal{V}^r$.
Again, this can only improve capacity.
Hence, these receivers are given $\mathcal{V}^r\cup\overline{\mathcal{V}}$, which consists of all the messages that receiver one should decode, as well as all the messages of all transmitters other than transmitter one.
Using this genie-given knowledge, every receiver can compute $\mathbf{x}_j$ for all $j\not=1$, and subtract all of them out of their output $\mathbf{y}_i$.
In other words, receiver $i\not=1$ can compute
\begin{equation}
\label{eq:phy-rxi-output}
\mathbf{y}_{i}'
= \mathbf{y}_i - \sum_{j\not=1} \mathbf{H}_{ij}\mathbf{x}_j
= \mathbf{H}_{i1}\mathbf{x}_1 + \mathbf{z}_i.
\end{equation}
Receiver $i$ is still expected to decode some messages.
Specifically, it must decode the subset of $\mathcal{V}^t$ that is intended for it, i.e.,
\[
\mathcal{V}^t_i = \left\{ V_{\mathcal{S}j} : 1\notin \mathcal{S} \text{ and } i\in \mathcal{S} \text{ and } j=1 \right\}.
\]
Then, by Fano's inequality,
\begin{equation}
\label{eq:phy-fano-i}
H\left( \mathcal{V}^t_i \middle| \mathbf{y}_i, \mathcal{V}^r, \overline{\mathcal{V}} \right)
\le \epsilon T.
\end{equation}

We focus now on receiver one.
From the problem requirements, it should be able to decode all of $\mathcal{V}^r$ with high probability.
After decoding $\mathcal{V}^r$, it has access to all the messages that receiver $i\not=1$ has, and hence it too can subtract out $\mathbf{x}_j$, $j\not=1$, from its output,
\[
\tilde{\mathbf{y}}_1'
= \tilde{\mathbf{y}}_1 - \sum_{j\not=1}\mathbf{H}_{1j}\mathbf{x}_j + \tilde{\mathbf{z}}_1
= \mathbf{H}_{11}\mathbf{x}_1 + \tilde{\mathbf{z}}_1.
\]
Since $\mathbf{H}_{11}$ is invertible almost surely, receiver one can then transform its output to get a statistical equivalent of the output of any other receiver.
Indeed, it can compute
\[
\breve{\mathbf{y}}_1^{(i)}
= \mathbf{H}_{i1}\mathbf{H}_{11}^{-1}\tilde{\mathbf{y}}_1'
= \mathbf{H}_{i1}\mathbf{x}_1 + \left( \mathbf{H}_{i1}\mathbf{H}_{11}^{-1} \right)\tilde{\mathbf{z}}_1
= \mathbf{H}_{i1}\mathbf{x}_1 + \breve{\mathbf{z}}_1^{(i)}.
\]
Since
$\breve{\mathbf{z}}_1^{(i)}=\left( \mathbf{H}_{i1}\mathbf{H}_{11}^{-1} \right)\tilde{\mathbf{z}}_1$
and the $\mathbf{H}_{ij}$ matrices are diagonal, then the variables $\breve{z}_1^{(i)}(\tau)$ are independent and have a variance of
\begin{IEEEeqnarray*}{rCl}
\var\left( \breve{z}_1^{(i)}(\tau) \right)
&=& \var\left( \frac{h_{i1}(\tau)}{h_{11}(\tau)} \tilde{z}_1(\tau) \right)\\
&=& \left( \frac{h_{i1}(\tau)}{h_{11}(\tau)} \right)^2 \cdot \var\left( \tilde{z}_1(\tau) \right)\\
&\le& 1,
\end{IEEEeqnarray*}
by \eqref{eq:phy-noise-variance}.
As a result, receiver one has at least as good a channel output as $\mathbf{y}_i'$ in \eqref{eq:phy-rxi-output}, and can thus decode anything that receiver $i$ can.
In particular, it can decode $\mathcal{V}^t_i$ for all $i$, i.e.,
\begin{equation}
\label{eq:phy-fano-1}
H\left( \mathcal{V}^t_i \middle| \tilde{\mathbf{y}}_1, \mathcal{V}^r, \overline{\mathcal{V}} \right)
\le H\left( \mathcal{V}^t_i \middle| \mathbf{y}_i, \mathcal{V}^r, \overline{\mathcal{V}} \right)
\le \epsilon T,
\end{equation}
using \eqref{eq:phy-fano-i}.

All of the above can be mathematically expressed in the following chain of inequalities, for any achievable $\tilde R_\sigma$.
\begin{IEEEeqnarray*}{rCl}
|\mathcal{V}| \cdot \tilde R_\sigma T
&=&
  H\left( \mathcal{V} \right)\\
&\overset{(a)}{=}&
  H\left( \mathcal{V} \middle| \overline{\mathcal{V}} \right)\\
&=&
  I\left( \mathcal{V} ; \tilde{\mathbf{y}}_1 \middle| \overline{\mathcal{V}} \right)
  + H\left( \mathcal{V} \middle| \tilde{\mathbf{y}}_1, \overline{\mathcal{V}} \right)\\
&=&
  I\left( \mathcal{V} ; \tilde{\mathbf{y}}_1 \middle| \overline{\mathcal{V}} \right)
  + H\left( \mathcal{V}^r \middle| \tilde{\mathbf{y}}_1, \overline{\mathcal{V}} \right)\\
&&{} + H\left( \mathcal{V}^t \middle| \tilde{\mathbf{y}}_1, \mathcal{V}^r, \overline{\mathcal{V}} \right)\\
&\overset{(b)}{\le}&
  I\left( \mathcal{V} ; \tilde{\mathbf{y}}_1 \middle| \overline{\mathcal{V}} \right)
  + \epsilon T
  + H\left( \mathcal{V}^t \middle| \tilde{\mathbf{y}}_1, \mathcal{V}^r, \overline{\mathcal{V}} \right)\\
&\overset{(c)}{\le}&
  I\left( \mathcal{V} ; \tilde{\mathbf{y}}_1 \middle| \overline{\mathcal{V}} \right)
  + \epsilon T
  + \sum_{i\not=1}
    H\left( \mathcal{V}^t_i \middle| \tilde{\mathbf{y}}_1, \mathcal{V}^r, \overline{\mathcal{V}} \right)\\
&\overset{(d)}{\le}&
  I\left( \mathcal{V} ; \tilde{\mathbf{y}}_1 \middle| \overline{\mathcal{V}} \right)
  + \epsilon T
  + \sum_{i\not=1} \epsilon T\\
&\overset{(e)}{\le}&
  I\left( \mathbf{x}_1,\ldots,\mathbf{x}_{K_t} ; \tilde{\mathbf{y}}_1 \middle| \overline{\mathcal{V}} \right)
  + K_r \epsilon T\\
&\overset{(f)}{\le}&
  T\cdot\left( \frac12\log\SNR + o(\log\SNR) \right) + K_r\epsilon T.
\end{IEEEeqnarray*}
In the above,
\begin{itemize}
\item $(a)$ is due to the independence of the messages;
\item $(b)$ uses Fano's inequality for receiver one;
\item $(c)$ follows from observing that $\mathcal{V}^t=\bigcup_{i\not=1} \mathcal{V}^t_i$;
\item $(d)$ uses \eqref{eq:phy-fano-1};
\item $(e)$ is due to the data processing inequality; and
\item $(f)$ is the MAC channel bound.
\end{itemize}
By taking $T\to\infty$ and $\epsilon\to0$, as well as $\SNR\to\infty$, we obtain
\[
\tilde d_\sigma^\star(K_t,K_r)
\le \frac{1}{|\mathcal{V}|}
= \frac{1}{|\mathcal{V}^r|+|\mathcal{V}^t|}
= \frac{1}{K_t\binom{K_r-1}{\sigma-1} + \binom{K_r-1}{\sigma}},
\]
thus proving \eqref{eq:dof-converse} and the converse part of Theorem~\ref{thm:phy}.
\hfill\IEEEQED

%% file: input_files/cadambe-lemmas.tex
In our interference alignment strategy, we use two crucial lemmas from \cite{cadambe2009}.
We present them here for ease of reference.

\begin{lemma}[from {\cite[Lemma~2]{cadambe2009}}]
\label{lemma:alignment}
Let $\mathbf{G}_1,\ldots,\mathbf{G}_\Gamma$ be $T\times T$ diagonal matrices, such that $G_g(\tau)$, the $\tau$-th diagonal entry of $\mathbf{G}_g$, follows a continuous distribution when conditioned on all other entries of all matrices.
Also let $\mathbf{b}$ be a column vector whose entries $b(\tau)$ are drawn iid from some continuous distribution, independently of $\mathbf{G}_1,\ldots,\mathbf{G}_\Gamma$.
Then, almost surely for any integer $n$ such that $T>(n+1)^\Gamma$, there exist matrices $\mathbf{A}_1$ and $\mathbf{A}_2$, of sizes $T\times(n+1)^\Gamma$ and $T\times n^\Gamma$ respectively, such that:
\begin{itemize}
\item
Every entry in the $\tau$-th row of $\mathbf{A}_1$ is a unique multi-variate monomial function of $b(\tau)$ and $G_g(\tau)$ for all $g$ ($b(\tau)$ and $G_g(\tau)$ appear with non-zero exponents in every entry), and the same is true for $\mathbf{A}_2$;%
\footnote{To clarify: a monomial could appear in both matrices $\mathbf{A}_1$ and $\mathbf{A}_2$, but never twice in the same matrix.}
and
\item
The matrices satisfy the following conditions almost surely,
\[
\mathbf{G}_g\mathbf{A}_2 \prec \mathbf{A}_1, \quad \forall g=1,\ldots,\Gamma,
\]
where $\mathbf{P}\prec\mathbf{Q}$ means that the span of the columns of $\mathbf{P}$ is a subspace of the space spanned by the columns of $\mathbf{Q}$.
\end{itemize}
\end{lemma}

\begin{lemma}[from {\cite[Lemma~1]{cadambe2009}}]
\label{lemma:decodability}
Let $x_i^{(k)}$, $i=1,\ldots,T$ and $k=1,\ldots,K$, be random variables such that each follows a continuous distribution when conditioned on all other variables.
Let $\mathbf{\Psi}$ be a $T\times T$ square matrix with entries $\psi_{ij}$ such that
\[
\psi_{ij} = \prod_{k=1}^K \left( x_i^{(k)} \right)^{p_{ij}^{(k)}},
\]
where $p_{ij}^{(k)}$ are integers such that
\[
\left( p_{ij}^{(1)},\ldots,p_{ij}^{(K)} \right)
\not=
\left( p_{ij'}^{(1)},\ldots,p_{ij'}^{(K)} \right),
\]
for all $i,j,j'$ such that $j\not=j'$.
In other words, the entries $\psi_{ij}$ are distinct monomials in the variables $x_i^{(k)}$.
Then, the matrix $\mathbf{\Psi}$ is almost surely full rank.
\end{lemma}

%% file: input_files/2x2-proofs.tex
Theorem~\ref{thm:2x2} gives an improved achievable DoF for the $2\times2$ cache-aided interference network.
In this appendix, we prove this result by describing and analyzing the interference-extraction scheme introduced in Section~\ref{sec:2x2} and illustrated in \figurename~\ref{fig:2x2-separation}, which achieves this DoF.

We describe the scheme in two steps.
First, we focus on the physical layer to show how more information can be extracted from the aligned interference at the receivers.
Second, we show how this additional information can be used at the network layer to achieve the inverse DoF in Theorem~\ref{thm:2x2}.

\subsection{Physical Layer}

In order to describe the interference-extraction scheme, let us first revisit the original separation architecture used when $M_r=0$.
The message set used for this case is the one where every transmitter has a message for every individual receiver, i.e., the unicast X-channel message set.
In order to achieve the optimal communication DoF of $1/3$ per message, at every receiver, the two messages intended for the other receiver are aligned in the same subspace.
Let us study this alignment more carefully.

Let $V_{ij}$ be the message intended for receiver $i$ from transmitter $j$.
Represent every message $V_{ij}$ by a scalar $v_{ij}$, called a stream.
By taking a block length of $3$ and by beamforming message $V_{ij}$ along some direction $\mathbf{a}_{ij}$, we get channel inputs
\[
\mathbf{x}_j = \mathbf{a}_{1j}v_{1j} + \mathbf{a}_{2j}v_{2j},
\]
and channel outputs
\begin{IEEEeqnarray*}{rCl}
\mathbf{y}_1 &=&
  \mathbf{H}_{11} \left(
      \mathbf{a}_{11}v_{11} + \mathbf{a}_{21}v_{21} \right)
  + \mathbf{H}_{12} \left(
      \mathbf{a}_{12}v_{12} + \mathbf{a}_{22}v_{22} \right)
  + \mathbf{z}_1;\\ \IEEEyesnumber\label{eq:yi-2x2}\IEEEyessubnumber\\
\mathbf{y}_2 &=&
  \mathbf{H}_{21} \left(
      \mathbf{a}_{11}v_{11} + \mathbf{a}_{21}v_{21} \right)
  + \mathbf{H}_{22} \left(
      \mathbf{a}_{12}v_{12} + \mathbf{a}_{22}v_{22} \right)
  + \mathbf{z}_2.\\ \IEEEyessubnumber
\end{IEEEeqnarray*}

\begin{lemma}
\label{lemma:yi-2x2}
We can choose the $\mathbf{a}_{ij}$'s in \eqref{eq:yi-2x2} such that
\begin{IEEEeqnarray*}{rCl}
\mathbf{y}_1 &=& \mathbf{\Psi}_1
\begin{bmatrix} v_{11} \\ v_{12} \\ v_{21}+v_{22} \end{bmatrix}
+ \mathbf{z}_1;\\
\mathbf{y}_2 &=& \mathbf{\Psi}_2
\begin{bmatrix} v_{21} \\ v_{22} \\ v_{11}+v_{12} \end{bmatrix}
+ \mathbf{z}_2,
\end{IEEEeqnarray*}
where the $3\times3$ matrices $\mathbf{\Psi}_i$ are full-rank almost surely.
\end{lemma}
\begin{IEEEproof}
Recall that the $\mathbf{H}_{ij}$'s are $3\times3$ diagonal matrices whose $\tau$-th diagonal element is $h_{ij}(\tau)$.
Also recall that these $h_{ij}(\tau)$ are independent and continuously distributed, which implies that $\mathbf{H}_{ij}$ is invertible almost surely.
Assume this invertibility is the case in the following.

Choose the $\mathbf{a}_{ij}$ vectors as:
\begin{IEEEeqnarray*}{rCl"rCl}
\mathbf{a}_{11} &=& \begin{bmatrix} 1 \\ 1 \\ 0 \end{bmatrix}; &
\mathbf{a}_{12} &=& \mathbf{H}_{22}^{-1}\mathbf{H}_{21}\mathbf{a}_{11};\\
\mathbf{a}_{21} &=& \begin{bmatrix} 1 \\ 0 \\ 1 \end{bmatrix}; &
\mathbf{a}_{22} &=& \mathbf{H}_{12}^{-1}\mathbf{H}_{11}\mathbf{a}_{21}.
\end{IEEEeqnarray*}

From \eqref{eq:yi-2x2}, the received signals are then
\begin{IEEEeqnarray*}{rCl}
\IEEEeqnarraymulticol{3}{l}{\mathbf{y}_1}\\
&=&
  \mathbf{H}_{11}\mathbf{a}_{11}v_{11}
  + \mathbf{H}_{12}\mathbf{H}_{22}^{-1}\mathbf{H}_{21}\mathbf{a}_{11}v_{12}
  + \mathbf{H}_{11}\mathbf{a}_{21} \left( v_{21} + v_{22} \right)\\
&&{} + \mathbf{z}_1\\
&=& \begin{bmatrix}
\mathbf{H}_{11}\mathbf{a}_{11} &
\mathbf{H}_{12}\mathbf{H}_{22}^{-1}\mathbf{H}_{21}\mathbf{a}_{11} &
\mathbf{H}_{11}\mathbf{a}_{21}
\end{bmatrix}
\begin{bmatrix} v_{11} \\ v_{12} \\ v_{21}+v_{22} \end{bmatrix}
+ \mathbf{z}_1\\
&=& \underbrace{\begin{bmatrix}
h_{11}(1) & \frac{h_{12}(1)h_{21}(1)}{h_{22}(1)} & h_{11}(1) \\
h_{11}(2) & \frac{h_{12}(2)h_{21}(2)}{h_{22}(2)} & 0 \\
0 & 0 & h_{11}(3)
\end{bmatrix}}_{\mathbf{\Psi}_1}
\begin{bmatrix} v_{11} \\ v_{12} \\ v_{21}+v_{22} \end{bmatrix}
+ \mathbf{z}_1,
\end{IEEEeqnarray*}
and, in a similar way,
\[
\mathbf{y}_2 = \underbrace{\begin{bmatrix}
h_{21}(1) & \frac{h_{22}(1)h_{11}(1)}{h_{12}(1)} & h_{21}(1) \\
0 & 0 & h_{21}(2) \\
h_{21}(3) & \frac{h_{22}(3)h_{11}(3)}{h_{12}(3)} & 0
\end{bmatrix}}_{\mathbf{\Psi}_2}
\begin{bmatrix} v_{21} \\ v_{22} \\ v_{11}+v_{12} \end{bmatrix}
+ \mathbf{z}_2.
\]
Since the $h_{ij}(\tau)$ are independent continuously distributed variables, then the matrices $\mathbf{\Psi}_1$ and $\mathbf{\Psi}_2$ are full-rank almost surely.
\end{IEEEproof}

Notice from Lemma~\ref{lemma:yi-2x2} that each receiver can recover, in addition to its intended streams, the sum of the streams intended for the other receiver.
By using a linear outer code over some finite field, we can ensure that obtaining the sum of two streams, e.g., $v_{21}+v_{22}$, yields the sum of the two corresponding messages, e.g., $V_{21}\oplus V_{22}$, where $\oplus$ indicates addition over the finite field.
For simplicity, we assume that this field is $\mathsf{GF}(2)$, although any finite field gives the same result.
In other words, receiver one can decode $V_{11}$, $V_{12}$, and $(V_{21}\oplus V_{22})$, and receiver two can decode $V_{21}$, $V_{22}$, and $(V_{11}\oplus V_{12})$.
Therefore, for the same per-message DoF of $1/3$, we get the linear combinations of the unintended messages for free.

\subsection{Network Layer}

\figurename~\ref{fig:2x2-separation} illustrates the interface between the physical and network layers resulting from the decoding of the aligned interference at each receiver.
This aligned interference, while available for free (no drawbacks in the communication DoF at the physical layer), becomes useful when the receiver memory is non-zero.
It provides a middle ground between pure unicast messages (as is done at $M_r=0$) and pure broadcast messages (which we use when $M_r=1$).

Let $\ell$ denote the link load, i.e., the size of each message $V_{ij}$, and let $L=4\ell$ be the sum network load.
For this specific separation architecture, we denote by $L^\star(M_r)$ the smallest sum network load as a function of receiver memory $M_r$, and by $\ell^\star=L^\star/4$ the smallest individual link load.
Since each message $V_{ij}$ (link) can be communicated across the physical layer using a DoF of $1/3$ by Lemma~\ref{lemma:yi-2x2}, then we can achieve an end-to-end DoF of
\begin{equation}
\label{eq:2x2-dof-load}
\frac{1}{\DoF} \le \frac{\ell^\star}{1/3} = \frac34 L^\star.
\end{equation}

Theorem~\ref{thm:2x2} follows directly from combining \eqref{eq:2x2-dof-load} with the following lemma.

\begin{lemma}
\label{lemma:2x2-achievability}
For the separation architecture illustrated in \figurename~\ref{fig:2x2-separation}, we can achieve the following sum network load:
\[
L^\star(M_r) \le \max\left\{
2 - 2M_r,
\frac{12}{7} - \frac87 M_r,
\frac43 - \frac23 M_r
\right\},
\]
for $M_r\in[0,2]$.
\end{lemma}

\begin{IEEEproof}
In order to prove Lemma~\ref{lemma:2x2-achievability}, it suffices to look at the following four $(M_r,L)$ corner points, as the rest can be achieved using time- and memory-sharing:
\[
\left( 0,2 \right),
\quad \left( 1/3, 4/3 \right),
\quad \left( 4/5, 4/5 \right),
\quad \left( 2,0 \right).
\]
The fourth corner point is trivial since $M_r=2$ implies each user can cache the entire library, and hence there is no need to transmit any information across the network.
The first corner point can be achieved by ignoring the aligned interference messages, which reduces to the original strategy.
Therefore, we only need to show the achievability of the second and third corner points.
For convenience, we will call the two files in the content library $A$ and $B$.

\paragraph*{Achieving point $(M_r,L)=(1/3,4/3)$}

When $M_r=1/3$, we split each file into three equal parts, labeled $A=(A_1,A_2,A_3)$ and $B=(B_1,B_2,B_3)$.

\begin{figure}
\centering
\includegraphics[scale=\myfigsscale]{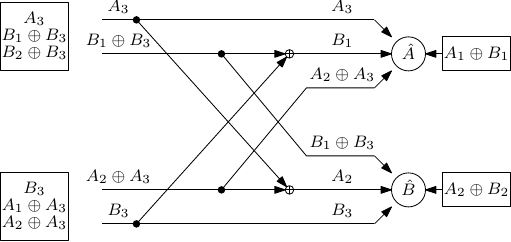}
\caption{Strategy for $M_r=1/3$, when the user requests are $(A,B)$.}
\label{fig:2x2-m13}
\end{figure}

\begin{table}
\centering
\caption{Achievable strategy for $M_r=1/3$.}
\label{tbl:delivery-m13}
\begin{tabular}{|c|c|c|c|c|c|}
\hline
Cache & \multicolumn{4}{c|}{Content} & Rx\\
\hline
Tx 1 & \multicolumn{4}{c|}{$A_3,B_1\oplus B_3,B_2\oplus B_3$} & N/A\\
Tx 2 & \multicolumn{4}{c|}{$B_3,A_1\oplus A_3,A_2\oplus A_3$} & N/A\\
\rowcolor{user1}
Rx 1 & \multicolumn{4}{>{\columncolor{user1}}c|}{$A_1\oplus B_1$} & 1\\
\rowcolor{user2}
Rx 2 & \multicolumn{4}{>{\columncolor{user2}}c|}{$A_2\oplus B_2$} & 2\\
\hline
\hline
& \multicolumn{4}{c|}{Demands $(\text{Rx1},\text{Rx2})$} &\\
Message & $(A,A)$ & $(A,B)$ & $(B,A)$ & $(B,B)$ & Rx\\
\hline
\rowcolor{user1}
$V_{11}$ & $A_3$ & $A_3$ & $B_2\oplus B_3$ & $B_1\oplus B_3$ & 1\\
\rowcolor{user2}
$V_{21}$ & $A_3$ & $B_1\oplus B_3$ & $A_3$ & $B_2\oplus B_3$ & 2\\
\rowcolor{user1}
$V_{12}$ & $A_1\oplus A_3$ & $A_2\oplus A_3$ & $B_3$ & $B_3$ & 1\\
\rowcolor{user2}
$V_{22}$ & $A_2\oplus A_3$ & $B_3$ & $A_1\oplus A_3$ & $B_3$ & 2\\
\hline
\rowcolor{user1}
$V_{21}\oplus V_{22}$ & $A_2$ & $B_1$ & $A_1$ & $B_2$ & 1\\
\rowcolor{user2}
$V_{11}\oplus V_{12}$ & $A_1$ & $A_2$ & $B_2$ & $B_1$ & 2\\
\hline
\end{tabular}
\end{table}

Table~\ref{tbl:delivery-m13} shows the placement and delivery phases, for all possible user requests, and \figurename~\ref{fig:2x2-m13} illustrates the strategy when the demands are $(A,B)$.
Notice that the transmitter caches hold exactly one file each (thus $M_t=1$), the receivers cache one third of a file each ($M_r=1/3$).
Furthermore, the messages $V_{ij}$ each carry the equivalent of one third of a file, which implies that $\ell=1/3$ is achieved, or, equivalently, a sum network load of $L=4\ell=4/3$.

\paragraph*{Achieving point $(M_r,L)=(4/5,4/5)$}

When $M_r=4/5$, we split each file into five equal parts, labeled $A=(A_1,\ldots,A_5)$ and $B=(B_1,\ldots,B_5)$.
For convenience, we define
\begin{IEEEeqnarray*}{rcl/rcl/rcl/rcl}
S_1 &=& B_2\oplus A_4, &
S_2 &=& A_1\oplus B_3, &
S_3 &=& B_1\oplus B_3, &
S_4 &=& B_2\oplus B_4,\\
T_1 &=& A_1\oplus A_3, &
T_2 &=& A_2\oplus A_4, &
T_3 &=& B_1\oplus A_3, &
T_4 &=& A_2\oplus B_4,
\end{IEEEeqnarray*}
and write $\mathcal{S}=\{S_1,S_2,S_3,S_4\}$ and $\mathcal{T}=\{T_1,T_2,T_3,T_4\}$.

\begin{figure}
\centering
\includegraphics[scale=\myfigsscale]{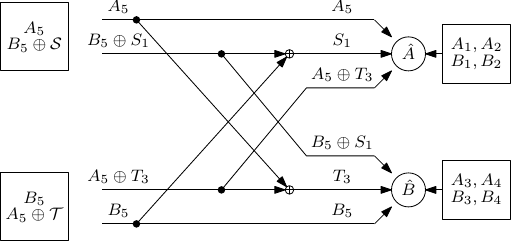}
\caption{Strategy for $M_r=4/5$, when the user requests are $(A,B)$.}
\label{fig:2x2-m45}
\end{figure}

\begin{table}
\centering
\caption{Achievable strategy for $M_r=4/5$.}
\label{tbl:delivery-m45}
\begin{tabular}{|c|c|c|c|c|c|}
\hline
Cache & \multicolumn{4}{c|}{Content} & Rx\\
\hline
Tx 1 & \multicolumn{4}{c|}{$A_5,B_5\oplus S_1,B_5\oplus S_2,B_5\oplus S_3,B_5\oplus S_4$} & N/A\\
Tx 2 & \multicolumn{4}{c|}{$B_5,A_5\oplus T_1,A_5\oplus T_2,A_5\oplus T_3,A_5\oplus T_4$} & N/A\\
\rowcolor{user1}
Rx 1 & \multicolumn{4}{>{\columncolor{user1}}c|}{$A_1,A_2,B_1,B_2$} & 1\\
\rowcolor{user2}
Rx 2 & \multicolumn{4}{>{\columncolor{user2}}c|}{$A_3,A_4,B_3,B_4$} & 2\\
\hline
\hline
& \multicolumn{4}{c|}{Demands $(\text{Rx1},\text{Rx2})$} &\\
Message & $(A,A)$ & $(A,B)$ & $(B,A)$ & $(B,B)$ & Rx\\
\hline
\rowcolor{user1}
$V_{11}$ & $A_5$ & $A_5$ & $B_5\oplus S_2$ & $B_5\oplus S_3$ & 1\\
\rowcolor{user2}
$V_{21}$ & $A_5$ & $B_5\oplus S_1$ & $A_5$ & $B_5\oplus S_4$ & 2\\
\rowcolor{user1}
$V_{12}$ & $A_5\oplus T_1$ & $A_5\oplus T_3$ & $B_5$ & $B_5$ & 1\\
\rowcolor{user2}
$V_{22}$ & $A_5\oplus T_2$ & $B_5$ & $A_5\oplus T_4$ & $B_5$ & 2\\
\hline
\rowcolor{user1}
$V_{21}\oplus V_{22}$ & $T_2$ & $S_1$ & $T_4$ & $S_4$ & 1\\
\rowcolor{user2}
$V_{11}\oplus V_{12}$ & $T_1$ & $T_3$ & $S_2$ & $S_3$ & 2\\
\hline
\end{tabular}
\end{table}

Table~\ref{tbl:delivery-m45} shows the placement and delivery phases, for all possible user requests, and \figurename~\ref{fig:2x2-m45} illustrates the strategy when the demands are $(A,B)$.
Notice that the transmitter caches hold exactly one file each (thus $M_t=1$), the receivers cache four fifths of a file each ($M_r=4/5$).
Furthermore, the messages $V_{ij}$ each carry the equivalent of one fifth of a file, which implies that $\ell=1/5$ is achieved, or, equivalently, a sum network load of $L=4\ell=4/5$.

By achieving all four corner points, we have proved Lemma~\ref{lemma:2x2-achievability}.
\end{IEEEproof}

\subsection{Optimality Within the Considered Separation Architecture}

Within the separation architecture considered throughout this appendix and Section~\ref{sec:2x2}, i.e., the one illustrated in \figurename~\ref{fig:2x2-separation}, we can show that the network-layer scheme is in fact exactly optimal.
Specifically, the sum network load achieved in Lemma~\ref{lemma:2x2-achievability} is optimal.
This is summarized in the following result.
\begin{proposition}
\label{prop:2x2-converse}
For all $M_r$, the optimal sum network load must satisfy
\[
L^\star(M_r) \ge \max\left\{
2 - 2M_r,
\frac{12}{7} - \frac87 M_r,
\frac43 - \frac23 M_r
\right\}.
\]
\end{proposition}
While this does not contribute to the main result in Theorem~\ref{thm:2x2}, it does reinforce it by showing that this is the best we can do within this separation architecture.

\begin{IEEEproof}
For the proof, it is more convenient to write the outer bounds in terms of the individual link load $\ell^\star=L^\star/4$.
Therefore, we will prove Proposition~\ref{prop:2x2-converse} by proving the following three inequalities (which together constitute an equivalent result):
\begin{IEEEeqnarray*}{rCrCl}
4\ell^\star &+& 2M_r &\ge& 2;\\
7\ell^\star &+& 2M_r &\ge& 3;\\
6\ell^\star &+&  M_r &\ge& 2.
\end{IEEEeqnarray*}

In the following, we refer to the two files as $A$ and $B$.
Let the cache contents of receivers one and two be $Q_1$ and $Q_2$, respectively.
We also write $V_{ij}^{ST}$ to denote the message $V_{ij}$ when user one has requested file $S$ and user two has requested file $T$, where $S,T\in\{A,B\}$.
Furthermore, we use $\mathcal{V}^{ST}$ to refer to all four messages when the requests are $S$ and $T$, and write $\mathcal{Y}_i^{ST}$ to denote the three outputs at receiver $i\in\{1,2\}$ when the requests are $S$ and $T$.
Therefore,
\begin{IEEEeqnarray*}{rCl}
\mathcal{V}^{ST} &=& \left( V_{11}^{ST}, V_{12}^{ST}, V_{21}^{ST}, V_{22}^{ST} \right);\\
\mathcal{Y}_1^{ST} &=& \left( V_{11}^{ST}, V_{12}^{ST}, V_{21}^{ST} \oplus V_{22}^{ST} \right);\\
\mathcal{Y}_2^{ST} &=& \left( V_{21}^{ST}, V_{22}^{ST}, V_{11}^{ST} \oplus V_{12}^{ST} \right).
\end{IEEEeqnarray*}

We will next prove each of the three inequalities.

\paragraph*{First inequality}
\begin{IEEEeqnarray*}{rCl}
(4\ell^\star + 2M_r)F
&\ge& H\left( Q_1,Q_2, \mathcal{V}^{AB} \right)\\
&=& H\left( Q_1,Q_2, \mathcal{V}^{AB} \middle| A,B \right)\\
&&{} + I\left( A,B ; Q_1,Q_2,\mathcal{V}^{AB} \right)\\
&=& H\left( Q_1,Q_2, \mathcal{V}^{AB} \middle| A,B \right)\\
&&{} + H\left(A,B\right) - H\left(A,B\middle|Q_1,Q_2,\mathcal{V}^{AB}\right)\\
&\overset{(a)}{\ge}& H\left(A,B\right) - \epsilon F\\
&=& 2F - \epsilon F,
\end{IEEEeqnarray*}
where $(a)$ is due to Fano's inequality.
Therefore,
\[
4\ell^\star + 2M_r \ge 2.
\]

\paragraph*{Second inequality}
\begin{IEEEeqnarray*}{rCl}
(7\ell^\star + 2M_r)F
&\ge& H\left( Q_1,\mathcal{Y}_1^{AB} \right)
+ H\left( Q_2,\mathcal{V}^{BA} \right)\\
&\overset{(a)}{\ge}& H\left( Q_1,\mathcal{Y}_1^{AB} \middle| A \right)
+ H\left( Q_2,\mathcal{V}^{BA} \middle| A \right)\\
&&{} + 2H\left(A\right) - 2\epsilon F\\
&\ge& H\left( Q_1,Q_2,\mathcal{Y}_1^{AB},\mathcal{V}^{BA} \middle| A \right)\\
&&{} + 2H\left(A\right) - 2\epsilon F\\
&\overset{(b)}{\ge}& H\left( Q_1,Q_2,\mathcal{Y}_1^{AB},\mathcal{V}^{BA} \middle| A,B \right)\\
&&{} + H\left(B\right) + 2H\left(A\right) - 3\epsilon F\\
&\ge& 3F - 3\epsilon F,
\end{IEEEeqnarray*}
where $(a)$ and $(b)$ once again follow from Fano's inequality.
Therefore,
\[
7\ell^\star + 2M_r \ge 3.
\]

\paragraph*{Third inequality}
\begin{IEEEeqnarray*}{rCl}
(6\ell^\star + M_r)F
&\ge& H\left( Q_1, \mathcal{Y}_1^{AA}, \mathcal{Y}_2^{BB} \right)\\
&\overset{(a)}{\ge}& H\left( Q_1,\mathcal{Y}_1^{AA},\mathcal{Y}_2^{BB} \middle| A,B \right)\\
&&{} + H\left(A,B\right) - \epsilon F\\
&\ge& 2F - \epsilon F,
\end{IEEEeqnarray*}
where $(a)$ is again due to Fano's inequality.
Therefore,
\[
6\ell^\star + M_r \ge 2.
\]

This concludes the proof of Proposition~\ref{prop:2x2-converse}.
\end{IEEEproof}